\documentclass{aa}
\usepackage[dvips]{graphicx}
\usepackage{natbib}
\usepackage{txfonts}
\usepackage{color}

\newcommand{\teff}{T_{\rm{eff}}}
\begin{document}
  \title{Three-dimensional hydrodynamical simulations of red giant stars: semi-global models for the interpretation of interferometric observations}
\titlerunning{Three-dimensional hydrodynamical simulations of red giant stars and interferometry}

  \author{A. Chiavassa
         \inst{1}
         \and
         R. Collet \inst{1}
         \and
         L. Casagrande \inst{1}
         \and
         M. Asplund \inst{1}
         }

  \offprints{A. Chiavassa}

   \institute{Max-Planck-Institut f\"{u}r Astrophysik, Karl-Schwarzschild-Str. 1, Postfach 1317, DÐ85741 Garching b. M\"{u}nchen, Germany\\
             \email{chiavass@mpa-garching.mpg.de}
                   }

  \date{Received; accepted }

 \abstract
{Theoretical predictions from models of red giant branch stars are a valuable
  tool for various applications in astrophysics ranging from galactic chemical evolution to studies of exoplanetary systems.}
{We use the radiative transfer code {{\sc Optim3D}} and realistic 3D
  radiative-hydrodynamical (RHD) surface convection simulations of red giants to explore the impact of granulation on interferometric observables.  We assess how 3D simulations of surface convection can
be validated against observations.}
{We compute intensity maps for the 3D simulation snapshots in two
  filters: in the optical at $5000\pm300$ \AA\
  and in the K band $2.14\pm0.26$ $\mu$m FLUOR filter,
  corresponding to the wavelength-range of instruments mounted on
   the CHARA interferometer. From the intensity maps, we construct
  images of the stellar disks, accounting for center-to-limb
  variations. We then derive interferometric visibility amplitudes
  and phases. We study their behavior with position angle and wavelength, and 
compare them with CHARA observations of the red giant star HD~214868.}
{We provide average limb-darkening coefficients for different metallicities and wavelength-ranges. We detail the prospects
for the detection and characterization of granulation and
center-to-limb variations of red giant stars with today's
interferometers. Regarding interferometric observables, we find
that the effect of 
	convective-related surface structures depends on metallicity and 
	surface gravity.  We provided theoretical closure phases that should be incorporated into the analysis of red giant planet companion closure phase signals. 	We estimate 
3D$-$1D corrections to stellar radii determination:
3D models are $\sim 3.5\%$ smaller to $\sim 1\%$ larger in the
optical with respect to 1D, and roughly $0.5$ to $1.5\%$ smaller in the infrared. Even if these corrections are
small, they are important to properly set the zero point of effective
temperature scale derived by interferometry and to strengthen the 
confidence of existing red giant catalogues of 	calibrating stars for interferometry. Finally, we show that our RHD
simulations provide an excellent fit to the red giant
HD~214868 even though more observations
are needed at higher spatial frequencies and shorter wavelength.}
{}

\keywords{      stars: horizontal-branch --
               stars: atmospheres --
               hydrodynamics --
               radiative transfer --
               techniques: interferometric 
               stars: individual: HD~214868
              }

  \maketitle

%

\section{Introduction}

Red giant branch stars have evolved from the main sequence and are powered by 
hydrogen burning in a thin shell surrounding their helium core. Their masses are typically less than $\sim$2.0
$M_\odot$ \citep{2002PASP..114..375S}, their effective temperatures
range from ${\sim}4\,000$ to ${\sim}5\,100$~K, depending on metallicity, their surface gravities from ${\sim}3.5$ to ${\sim}1.0$ in $\log{g}$
\citep{2007A&A...475.1003H}, and their radii from ${\sim}$3 to
${\sim}$70 R$_\odot$ \citep{1999AJ....117..521V,
  2010ApJ...710.1365B}. The determination of fundamental parameters
of red giant stars is of great importance in astrophysics: (i) red giant stars
will be used as tracers of the
 morphology and evolution of the Galaxy in the framework of the {\sc Gaia} mission
 \citep{2001A&A...369..339P, 2008IAUS..248..217L}, as they are
 intrinsically bright and thus can probe regions obscured by interstellar extinction; (ii) they
 are extensively used for spectroscopic elemental abundance analyses
 of distant stellar populations \citep[e.g.
 ][]{1995AJ....109.2757M, 1996ApJ...471..254R, 2000AJ....120.1841F};
 (iii) they are relatively easy targets in open
 \citep{2005AJ....130..597Y, 2005AJ....130.1111C} and globular
 clusters \citep[see ][for a review]{2004ARA&A..42..385G} and can 
	     be used for a number of purposes, including measuring Galactic 
	     metallicity gradients; (iv) red giant stars have been included in planet
search surveys \citep{2005ApJ...633..465S,2007A&A...472..649D,2008ApJ...675..784J}.

Three-dimensional hydrodynamical modeling of convection of stellar
surfaces has been already extensively employed to study the effects of
photospheric inhomogeneities and velocity fields on the formation of
spectral lines in a number of cases, including the Sun, dwarfs and
subgiants \citep[e.g. ][]{1999A&A...346L..17A, 2001A&A...372..601A,
 2009ARA&A..47..481A,2010SoPh..tmp...66C,2010A&A...513A..72B,2010arXiv1003.4510S}
and red giants \citep[e.g.][]{2007A&A...469..687C,
 2009MmSAI..80..719C,2009A&A...508.1429W,2009MmSAI..80..723K}. Such
simulations are paramount for an accurate quantitative analysis of observed data. These studies provide valuable information to properly understand astrophysical processes such as stellar nucleosynthesis, internal mixing mechanisms in stars, and Galactic chemical evolution.

Within this framework, the correct interpretation of interferometric 
observables is very important to properly 
recover fundamental stellar parameters via a correct description 
of limb-darkening laws. Once angular diameters are known, 
the combination with bolometric fluxes and parallaxes provides
effective temperatures ($\teff$) and physical radii. 
A proper chracterization of these quantities is also relevant 
from the point of view of stellar evolution, since red giants are 
populating the coolest and most luminous part 
of the Hertzsprung-Russell (HR) diagram.

In this work, we present interferometric predictions obtained from three-dimensional surface convection simulations of red giant stars with different effective temperatures, surface gravities, and metallicities.

\section{Surface convection simulations and radiative transfer calculations}

\subsection{Radiative-hydrodynamical simulations of red giant stars}\label{Sectmodel}

We adopt here time-dependent, three-dimensional (3D), radiative-hydrodynamical (RHD) surface convection simulations 
of red giant stars generated by \cite{2007A&A...469..687C} and \cite{2009MmSAI..80..719C} 
with the \cite{1998ApJ...499..914S} code and with the 
{\sc Stagger-Code},\footnote{www.astro.ku.dk/{\textasciitilde}aake/papers/95.ps.gz} respectively.
In these simulations, the equations of conservation of mass, momentum, and energy are solved
for representative rectangular volumes located across the stars' optical surfaces.
Heat losses and gains due to radiation are accounted for by solving the 3D radiative transfer equation in the
local thermodynamic equilibrium (LTE) approximation.
The simulations employ realistic input physics: the equation of state
is an updated version of the one by \cite{1988ApJ...331..815M}. Continuous and line opacities are taken from
\cite{1975A&A....42..407G,1992RMxAA..23..181K,1993KurCD..12.....K}.
The simulation domains are chosen large enough to cover at least ten pressure scale heights 
vertically and to allow for about ten granules to develop at the surface; 
they are periodic horizontally, while their top and bottom boundaries are left open.
The parameters of the adopted simulations are given in Table~\ref{simus}.
For all simulations, we assume a solar chemical composition from \cite{1998SSRv...85..161G} but with all metal abundances
scaled proportionally to the desired iron abundance. 
The simulations assume a constant gravitational acceleration parallel
to the vertical axis; hence, there is a degeneracy between stellar
mass and radius ($g\,{\propto}\,M/R^2$). Nonetheless, in the present work,
we assume for each simulation a mass of 0.8~M$_\odot$ -- typical of
red giants with stellar parameters like the ones considered here -- to
estimate the spatial frequency scale in
Section~\ref{SectVis}.

\begin{table*}
\begin{minipage}[t]{\textwidth}
\caption{Simulations of red giant stars used in this work. Note that the mass has been fixed to determine the radius.}             
\centering
\label{simus}      
\renewcommand{\footnoterule}{} 
\begin{tabular}{c c c c c c c}        
\hline\hline                 
$<T_{\rm{eff}}>$\footnote{Temporal average and standard deviation of the emergent effective temperatures} & [Fe/H]  & $\log g$ & $x,y,z$-dimensions & $x,y,z$-dimensions   &$M$ & $\rm{R}_{\star}$ \\
$[\rm{K}]$ & & [cgs]  & [Mm]  & [grid points]   & [$M_\odot$] & [$\rm{R}_\odot$]\\
\hline
4697$\pm$18\footnote{\cite{2007A&A...469..687C}} & 0.0 & 2.2 &  1250$\times$1250$\times$610  & 100$\times$100$\times$125 & 0.8 & 12.9 \\
4717$\pm$12$^b$ & $-$1.0 & 2.2 & 1125$\times$1125$\times$415 &100$\times$100$\times$125  &0.8 & 12.9\\
5035$\pm$13$^b$ & $-$2.0  & 2.2 & 1150$\times$1150$\times$430 &100$\times$100$\times$125  &0.8 & 12.9\\
5128$\pm$10$^b$ & $-$3.0  & 2.2 & 1150$\times$1150$\times$430 & 100$\times$100$\times$125  & 0.8 & 12.9\\
4627$\pm$14\footnote{\cite{2009MmSAI..80..719C} and Collet et al. 2010, in prep.}  & $-$3.0 &1.6 & 3700$\times$3700$\times$1100  & 480$\times$480$\times$240 & 0.8 & 23.0\\
\hline\hline                          
\end{tabular}
\end{minipage}
\end{table*}

\subsection{Post processing radiative transfer code}

\begin{figure}
\centering
     \includegraphics[width=85mm]{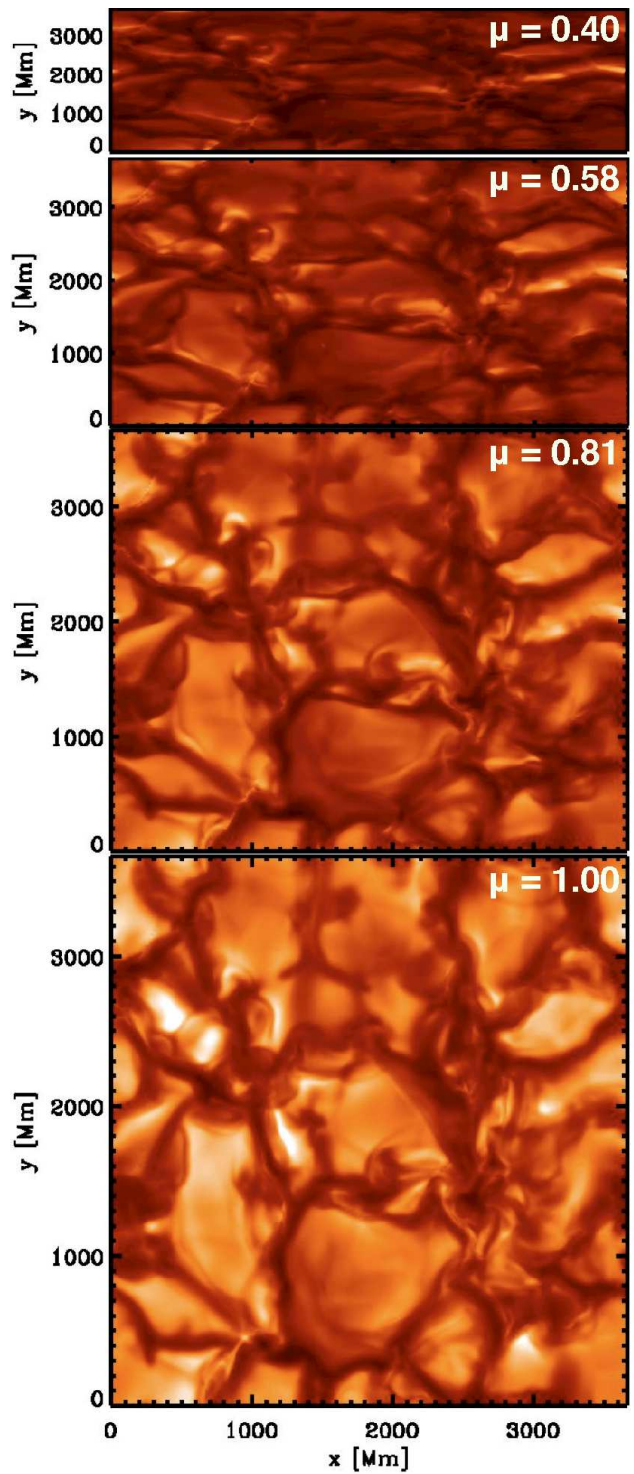} 
     \caption{Intensity maps in the optical top-hat filter of Fig.~\ref{filters} for the simulation with $\log g=1.6$ and [Fe/H]=$-$3.0 (Table~\ref{simus}) with $\mu$=1.00, 0.81, 0.58, and 0.40. The intensities typically ranges from $3{\times}10^4$ to $1.2\times10^6$\,erg\,cm$^{-2}$\,s$^{-1}$\,{\AA}$^{-1}$.
          }
       \label{limb}
  \end{figure}

We used the 3D pure-LTE radiative transfer code {\sc Optim3D}
\citep{2009A&A...506.1351C} to compute intensity maps from
the snapshots of the RHD simulations listed in Table~\ref{simus}. The code takes into account the
Doppler shifts due to convective motions. The radiative
transfer equation is solved monochromatically using extinction
pre-tabulated coefficients as a function of temperature, density, and
wavelength. The lookup tables were computed for the same chemical compositions as the RHD simulations
using the same extensive atomic and molecular opacity data as the latest generation of
MARCS models \citep{2008A&A...486..951G}. We assumed a zero
micro-turbulence since the velocity fields inherent in 3D models
are expected to self-consistently and adequately account for
non-thermal Doppler broadening of spectral lines
\citep{2000A&A...359..755A}. The temperature and density ranges
spanned by the tables are optimized for the values encountered in the RHD simulations. \\
{\sc Optim3D} computes the emerging intensities for vertical rays cast through the
computational box, for all required wavelengths, with the
method described by \cite{2009A&A...506.1351C}. The procedure is
repeated after tilting the computational box by an angle $\theta$ with respect to
the line of sight (vertical axis) and rotating it azimuthally by an angle $\phi$. 
The final result is a spatially resolved intensity spectrum at different
angles. This implementation of {\sc Optim3D} enables radiative
transfer calculations with RHD simulations either in the \emph{star-in-a-box} configuration \cite[e.g.][]{2009A&A...506.1351C} or in the \emph{box-in-a-star} configuration (this work).

\section{Sphere tiling: from local images to a semi-global intensity maps}\label{Sectionortho}

In this Section, we aim to obtain an image of the stellar disk as a
nearby observer would see it as required to extract the
interferometric observables.

The computational domain of each simulation represents only a small portion of the stellar surface (see Section~\ref{Sectmodel}). 
To overcome this limitation, and at the same time account for limb-darkening effects, we computed surface intensity maps for different inclinations with respect to the vertical ($\theta$-angles) and for representative series of simulation snapshots\footnote{For each simulation, we typically selected ${\sim}30$ snapshots, taken at regular intervals and covering ${\sim}100-250$ hours of stellar time} and used them to tile a spherical surface.
The actual value of the $\theta$-angle used to generate each map depended on the position (longitude and latitude) of the tile on the sphere and was determined by computing the angular distance of the tile from the center of the stellar disk ($0^{\circ}$ in longitude and latitude on the sphere).
For practical reasons, we actually computed intensity maps for a set of 15 predefined values of $\mu{\equiv}\cos(\theta)$=[1.000, 0.989, 0.978, 0.946, 0.913, 0.861, 0.809, 0.739, 0.669, 0.584, 0.500, 0.404, 0.309, 0.206, 0.104] (Fig.~\ref{limb}) and interpolated linearly in between them.
Mapping onto a spherical surface induces distortions especially at high latitudes and longitudes; we accounted for that by appropriately cropping the square-shaped intensity maps when defining the spherical tiles.

\begin{figure}[!h]
  \centering
     \includegraphics[width=0.9\hsize]{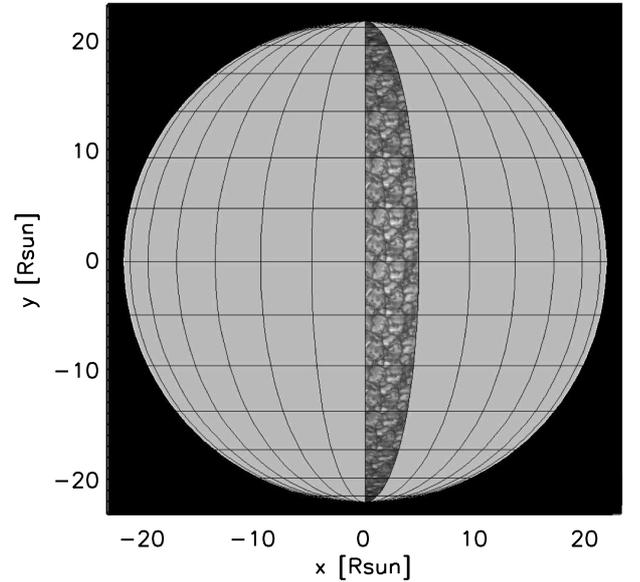} 
     \caption{Toy model representing our method used to tile spherical
surfaces and constructing the synthetic stellar disk images.
          }
       \label{toy}
  \end{figure}

In addition, we took into consideration statistical tile-to-tile fluctuations in the number of granules and in their shape and size and computed intensity maps by selecting snapshots at random from each simulation's time-series. 
This random selection process also avoided that the assumption of periodic boundary conditions resulted in a tiled spherical surface globally displaying an artifactual periodic granulation pattern.
Based on the stellar radii estimates given in Table~\ref{simus} and on the sizes of the simulations' domains, we required 14 to 23 tiles to
cover half a great circle from side to side on the sphere, depending on the surface gravity of the models ($\log{g}$=1.6 and 2.2, respectively).
The adopted number of $\mu$-angles for the pre-computed intensity maps is therefore sufficiently large to ensure an accurate representation of the center-to-limb variations for the present work. 
To produce the final stellar disk images, we performed an orthographic projection of the tiled spheres on a plane perpendicular to the line-of-sight ($\theta=0^{\circ}$). 
The orthographic projection returned images of the globes in which distortions are greatest toward the rim of the hemisphere where distances are compressed. The sketch in Fig.~\ref{toy} outlines our method we used to tile spherical surfaces and constructing the synthetic stellar disk images.

\begin{figure*}
  \centering
   \begin{tabular}{c}
     \includegraphics[width=0.5\hsize]{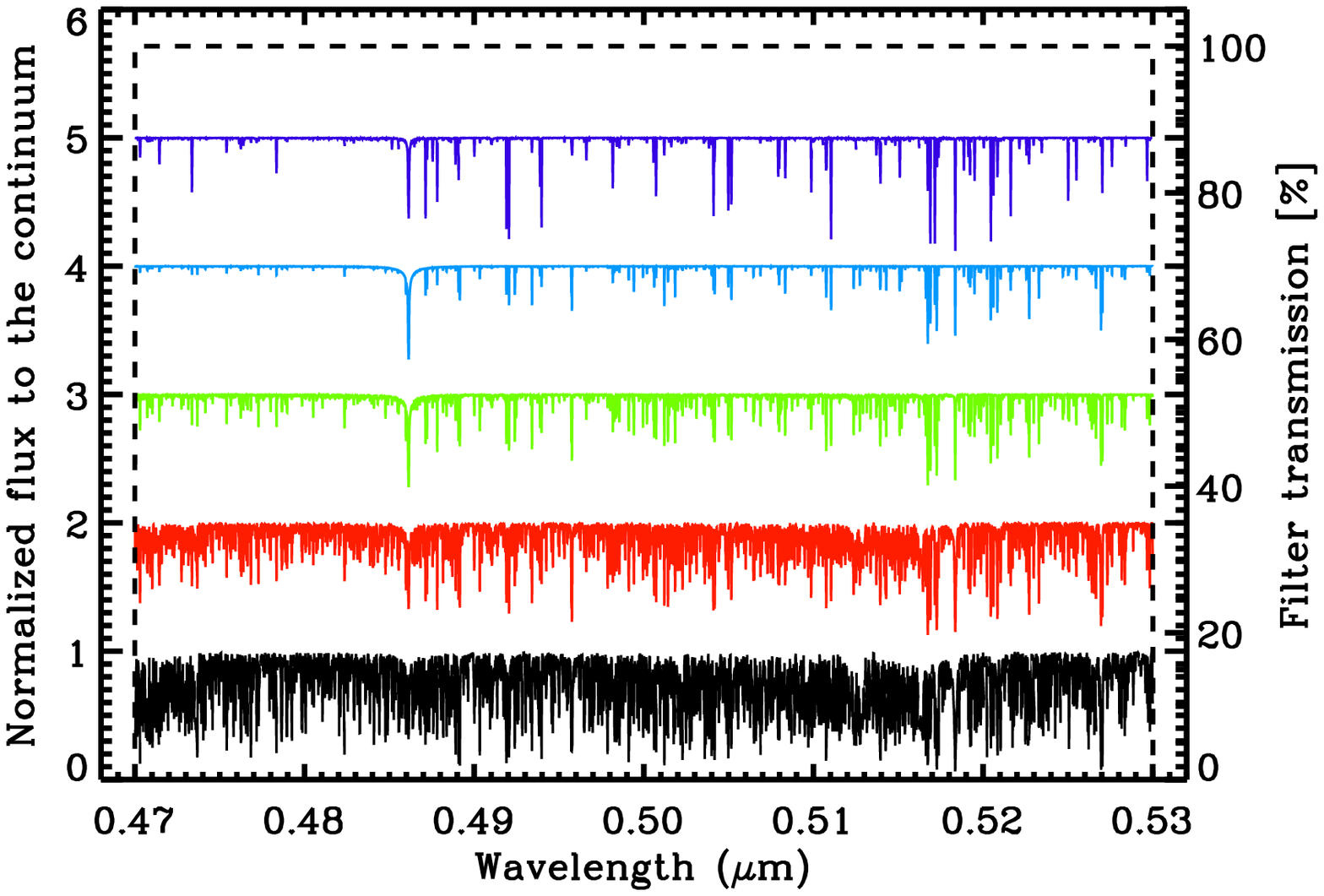} 
 \includegraphics[width=0.5\hsize]{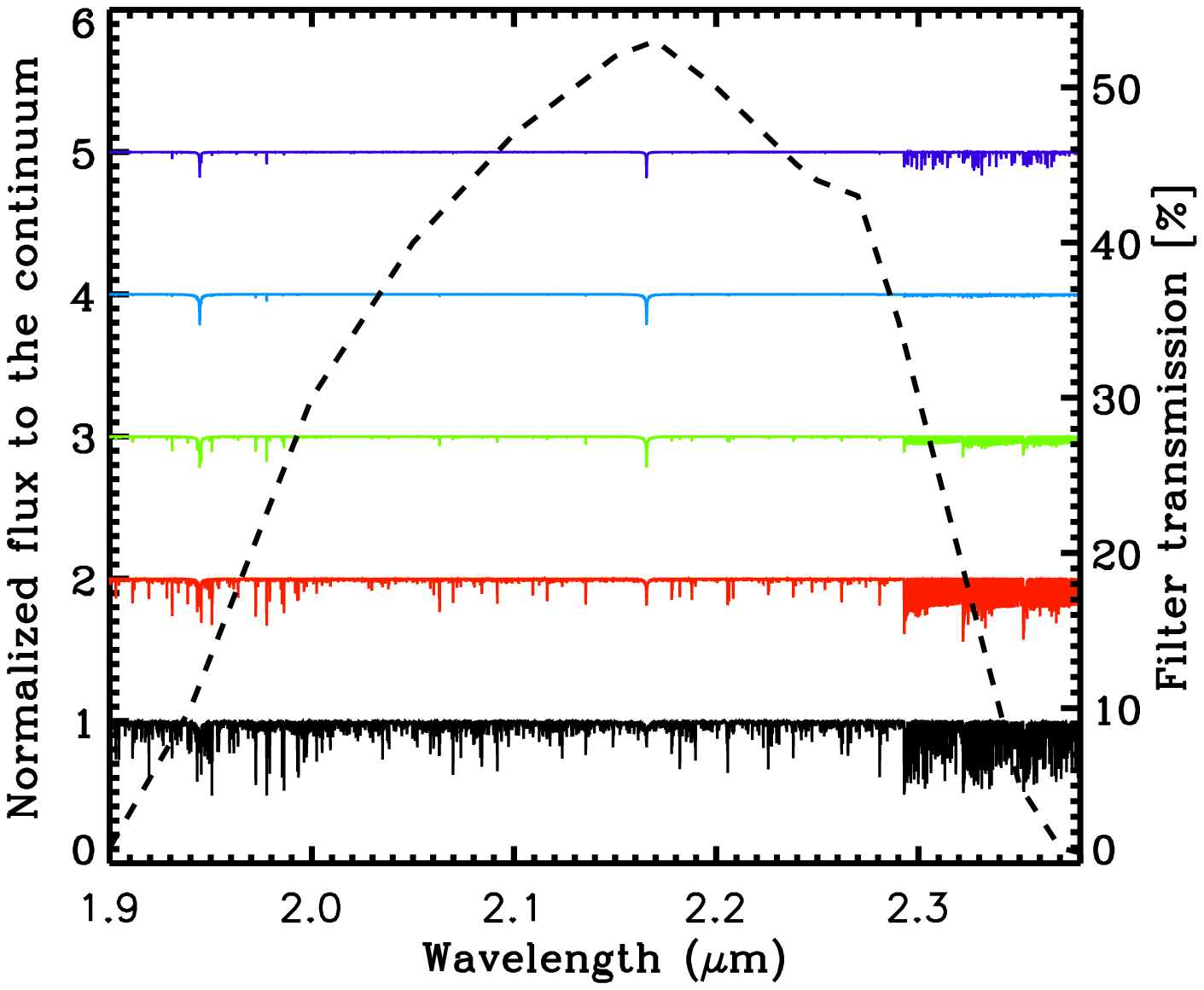} 
 \end{tabular}
     \caption{\emph{Left panel:} synthetic spectra for different 3D model red giant atmospheres in the wavelength range  $4700$--$5300$\,
       \AA\ . The intensity maps corresponding to this filter have
       been computed assuming
       a top-hat filter (dashed line). This wavelength range corresponds to the VEGA
       instrument
       \citep{2009A&A...508.1073M} mounted on CHARA with a spectral
       resolution $\lambda/\Delta\lambda=5000$. From the bottom: the black curve refers to
       the simulation with $\log g=2.2$ and [Fe/H]=0.0 and no
       offset applied, the red curve to $\log g=2.2$ and [Fe/H]=$-$1.0
       with an offset of 1, the green curve to $\log g=2.2$ and
       [Fe/H]=$-$2.0 with an offset of 2, the light blue curve to $\log
       g=2.2$ and [Fe/H]=$-$3.0 with an offset of 3, and the dark blue
       curve $\log g=1.6$ and [Fe/H]=$-$3.0 with an offset of 4 (see
       Table~\ref{simus}). \emph{Right panel:} transmission curve (dashed line)
       of the FLUOR ($2.14\pm0.26$ $\mu$m) filter
       \citep{2006SPIE.6268E..46M} mounted on CHARA. Colors and
        offsets are the same as in left panel.
          }
       \label{filters}
  \end{figure*}

For the computation of the interferometric observables, we consider
two spectral bands: (i) in the optical between 4700 and 5300 \AA\ and
(ii) in the infrared at 2.14$\pm$0.26 $\mu$m FLUOR filter,
Fig.~\ref{filters}. These wavelength intervals have been chosen because
they correspond to the range of instruments mounted on CHARA
interferometer \citep{2005ApJ...628..453T}. These telescopes are 
particularly suited for observations of red giant stars because of
its long baselines allow to resolve objects with small apparent diameters. The resulting intensity maps reported are normalized to the filter transmission as: 
$\frac{\int I_{\lambda} T\left(\lambda\right)d\lambda}{\int T\left(\lambda\right)d\lambda}$ where 
$I_\lambda$ is the intensity and $T\left(\lambda\right)$ is the transmission curve of the filter 
at a certain wavelength. The spectra in Fig.~\ref{filters} have been
computed along rays of four $\mu$-angles [0.88, 0.65, 0.55, 0.34] and four $\phi$-angles [$0^{\circ}$,
$90^{\circ}$, $180^{\circ}$, $270^{\circ}$], after which we
performed a disk integration and a temporal average over all selected
snapshots.

\section{Three-dimensional limb darkening coefficients}\label{Sectlimb}

\begin{figure*}
  \centering
   \begin{tabular}{c}
     \includegraphics[width=0.4\hsize]{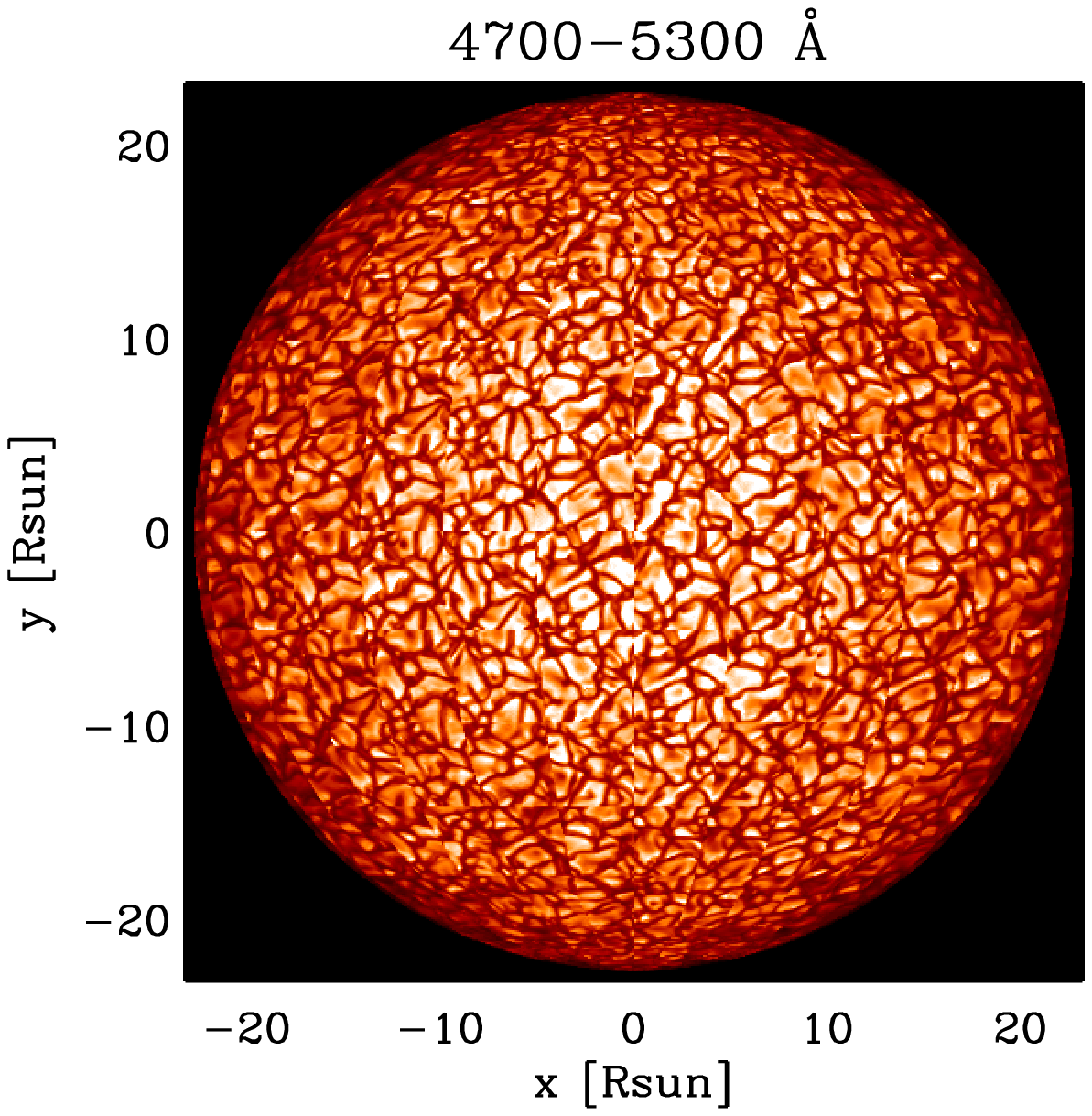} 
 \includegraphics[width=0.4\hsize]{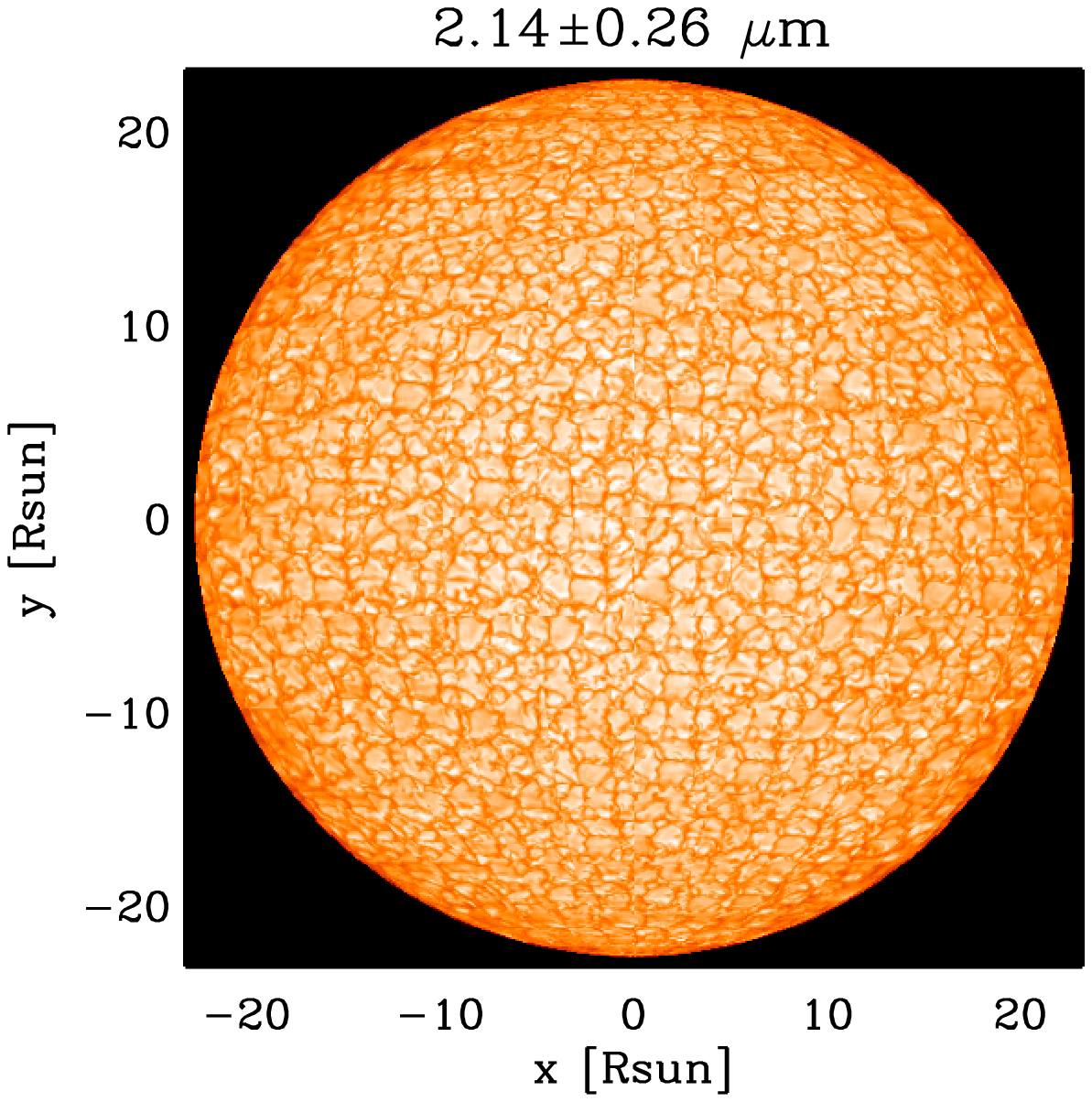} 
 \end{tabular}
     \caption{Synthetic stellar disk images of the simulation with [Fe/H]=$-$3.0 and $\log g$=1.6 (Table~\ref{simus}). The intensity range is $5.0{\times}10^3$--$2.1\times10^6$\,erg\,cm$^{-2}$\,s$^{-1}$\,{\AA}$^{-1}$ for the optical filter and
$5.0{\times}10^3$--$1.1\times10^5$\,erg\,cm$^{-2}$\,s$^{-1}$\,{\AA}$^{-1}$ for the FLUOR filter.
          }
       \label{intensity_images}
  \end{figure*}

In this Section we derive average center-to-limb intensity profiles
with their limb darkening (LD) laws for the simulations in
Table~\ref{simus}. Figure~\ref{intensity_images} shows the resulting synthetic stellar disk images for the [Fe/H]=$-$3.0 and $\log
g$=1.6 simulation in both filters of
Fig.~\ref{filters}. The surface of the simulated giants stars appear
irregular, permeated with structures. The center-to-limb
variations are visible in both images, but are more pronounced in the optical filter. This is due to the integrated extinction by atomic and
molecular lines in the upper layers which is more important in the
optical range than in the FLUOR filter: this difference further
increases for metal-poor models (see Fig.~\ref{filters}).

\begin{figure*}
  \centering
   \begin{tabular}{c}
     \includegraphics[width=0.5\hsize]{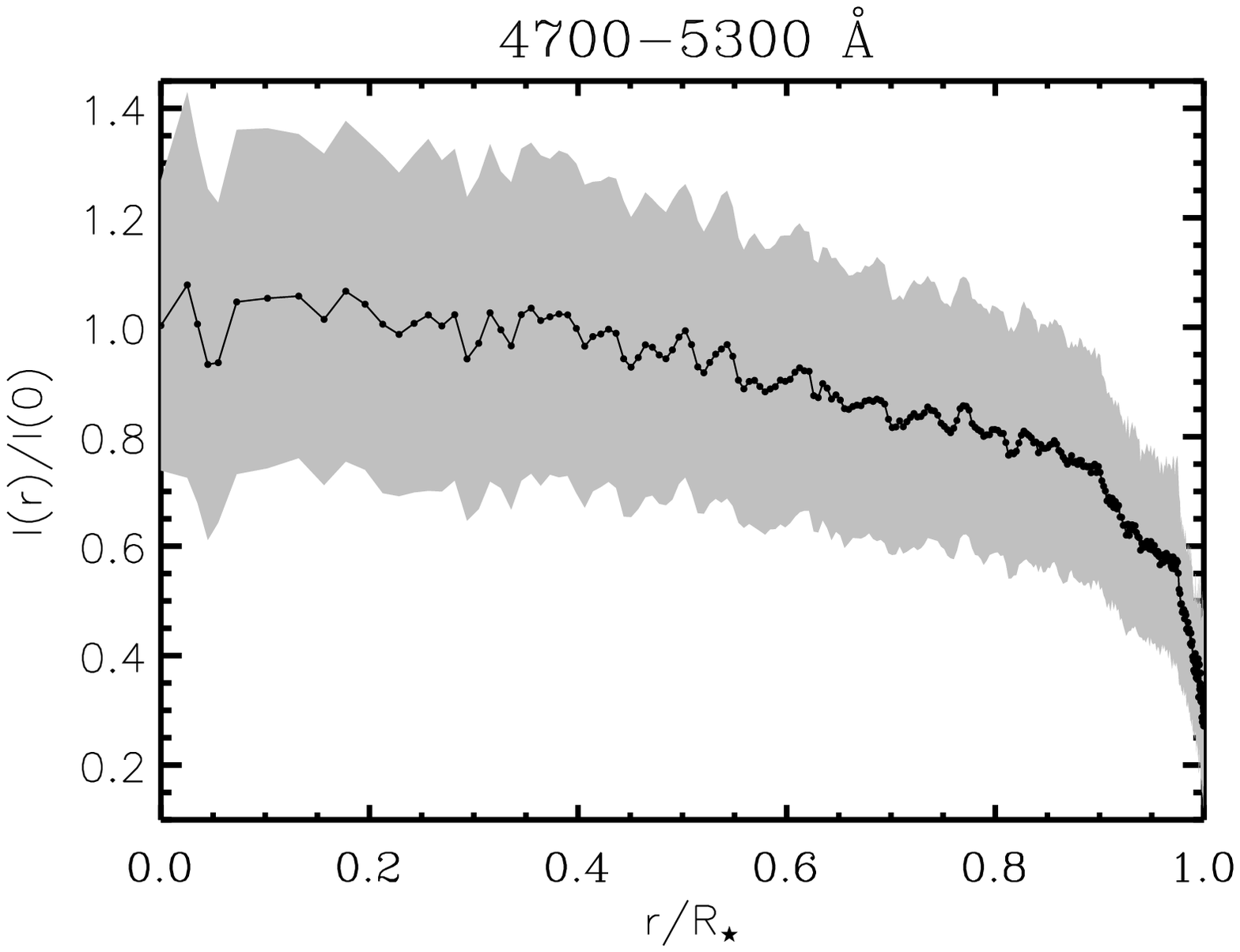} 
 \includegraphics[width=0.5\hsize]{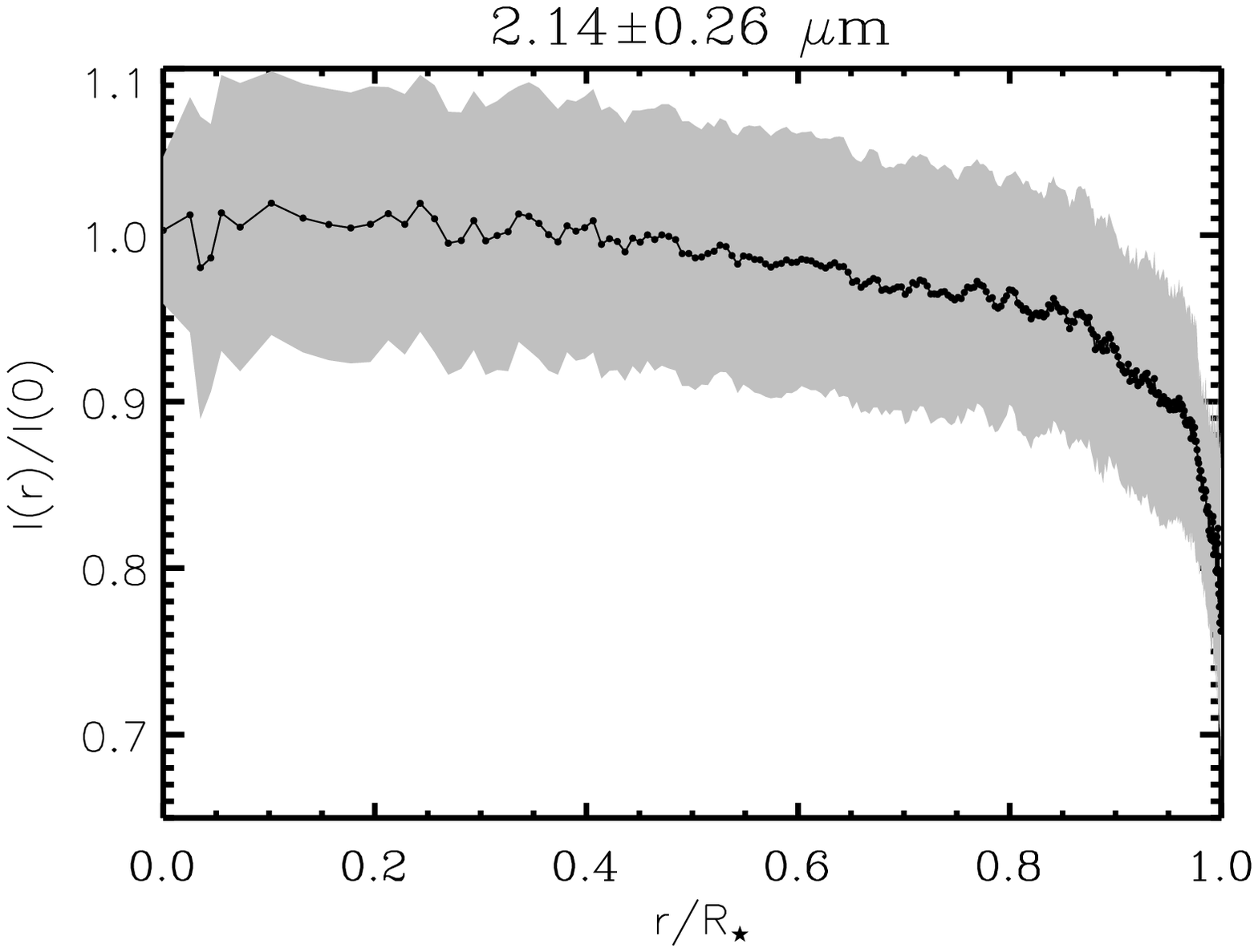} 
 \end{tabular}
     \caption{Radially averaged intensity profiles (black line)
       derived from the synthetic stellar disk images of
       Fig.~\ref{intensity_images} and corresponding to RHD
       simulation with [Fe/H]=$-$3.0 and $log g$=1.6. The dashed grey areas denote the
       one sigma spatial fluctuations with respect to the averaged
       intensity profile. The
       intensity is normalized to the mean intensity at disk center and the radius is normalized to the radius given in Table~\ref{simus}.
     }
       \label{intensity_profiles}
  \end{figure*}

We then derived azimuthally averaged intensity profiles for every
synthetic stellar disk image from the simulations, with two examples shown in
Fig.~\ref{intensity_profiles}. The profiles were constructed using
rings regularly spaced in $\mu=cos(\theta)$ for $\mu\le1$ (i.e.
$r/\rm{R}_{\star}\le1$), with $\theta$ the angle between the line of sight
and the radial direction. The parameter $\mu$ is related to the impact parameter
$r/\rm{R}_{\star}$ through $r/\rm{R}_{\star}=\sqrt{1-\mu^2}$, where $\rm{R}_{\star}$
is the stellar radius reported in Table~\ref{simus}. We ensure a good
characterization of the intensity towards the limb using 285
rings, which is half size in pixels of the images of Fig.~\ref{intensity_images}. The standard deviation of the average intensity,
$\sigma_{I\left(\mu\right)}$, was computed within each ring. As for
the images in Fig.~\ref{intensity_images}, the center-to-limb
variations is steeper in the case of the optical filter. Moreover, the fluctuations are stronger in the optical case with values up to 20$\%$ with respect to the average (Fig.~\ref{intensity_profiles}). 

We used these averaged profiles to determine LD laws for the simulations. 
For this purpose, we fitted the azimuthally averaged intensities determined from the simulations with a polynomial LD law of the form
\begin{equation}\label{claret_law}
\frac{I(\mu)}{I(1)}=\sum_{k=0}^N a_k\left(1-\mu\right)^k
\end{equation}
as in \cite{2009A&A...506.1351C}, where $I(\mu)$ is the intensity,
$a_k$ the LD coefficients, and $N+1$ their number. We
fitted all the azimuthally averaged profiles using this law and weighting the fit by $1/\sigma_{I\left(\mu\right)}$. We varied the order $N$ of the polynomial
LD laws and found that $N=2$ in practice already gave an optimal fit,
with the $N=3$ solution providing only minor and altogether negligible
improvements to the $\chi^2$ minimization. At the same time, we would like to stress that using
fewer terms in Eq.~\ref{claret_law} (e.g. a simple power law) does not provide a
good representation of the azimuthally averaged profiles and we therefore
discourage this practice. In Table~\ref{table1} we listed the LD coefficients $a_k$ for each simulation and for both filters considered in the present work.

\begin{table}
\begin{minipage}[t]{\columnwidth}
\caption{Limb-darkening coefficients (see Eq.~\ref{claret_law} with $N=2$) for the RHD simulations of Table~\ref{simus} at different wavelength filters. }
\label{table1}
\centering
\renewcommand{\footnoterule}{}  
\begin{tabular}{c|cc|ccc}
\hline \hline
$\lambda$ [$\mu$m] &  [Fe/H]   &  ${\rm log}~g$  &$a_{0}$     & $a_{1}$  &  $a_{2}$  \\
\hline
0.5 \footnote{central wavelength of the corresponding optical filter} & 0.0 &  2.2  & 1.000  & $-$0.912 & 0.150\\
2.14 \footnote{central wavelength of the corresponding FLUOR filter}& 0.0 & 2.2 & 1.000 & $-$0.156 & $-$0.167\\
\hline
0.5 & $-$1.0  &  2.2  & 1.000  & $-$0.846 & 0.080 \\
2.14& $-$1.0 & 2.2 & 1.000 & $-$0.146 & $-$0.148 \\
\hline
0.5 & $-$2.0  &  2.2  & 1.000  & $-$0.556 & $-$0.159\\
2.14& $-$2.0 & 2.2 & 1.000 & $-$0.108 & $-$0.13 \\
\hline
0.5 & $-$3.0  &  2.2  & 1.000  & $-$0.537 & $-$0.136 \\
2.14& $-$3.0 & 2.2 & 1.000 & $-$0.117 & $-$0.090 \\
\hline
0.5 & $-$3.0  &  1.6  & 1.000  & $-$0.316 & $-$0.410\\
2.14& $-$3.0 & 1.6 & 1.000 & 0.010 & $-$0.235 \\
\hline
\end{tabular}
\end{minipage}
\end{table}

\begin{figure*}
  \centering
   \begin{tabular}{cc}
 \includegraphics[width=0.5\hsize]{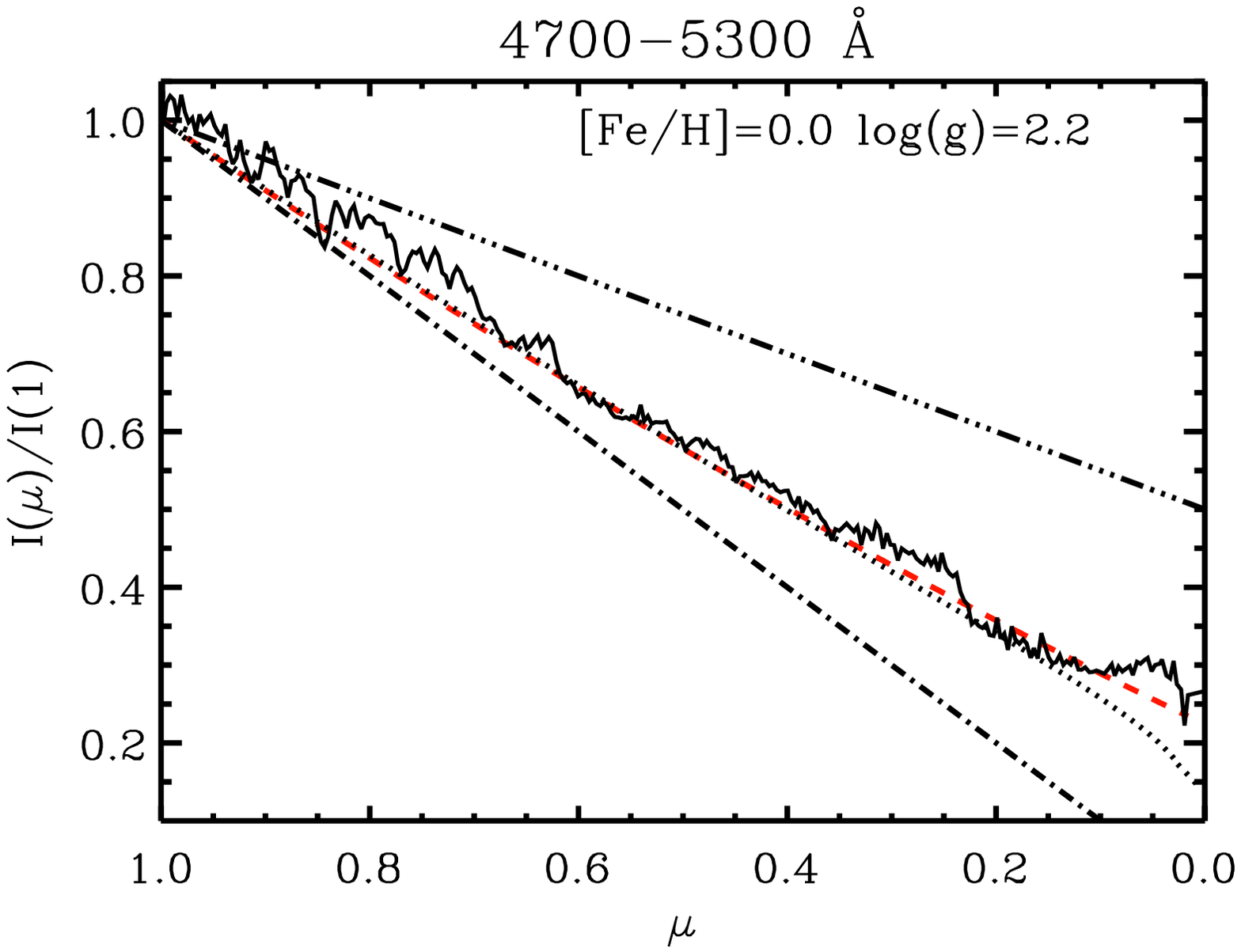} 
 \includegraphics[width=0.5\hsize]{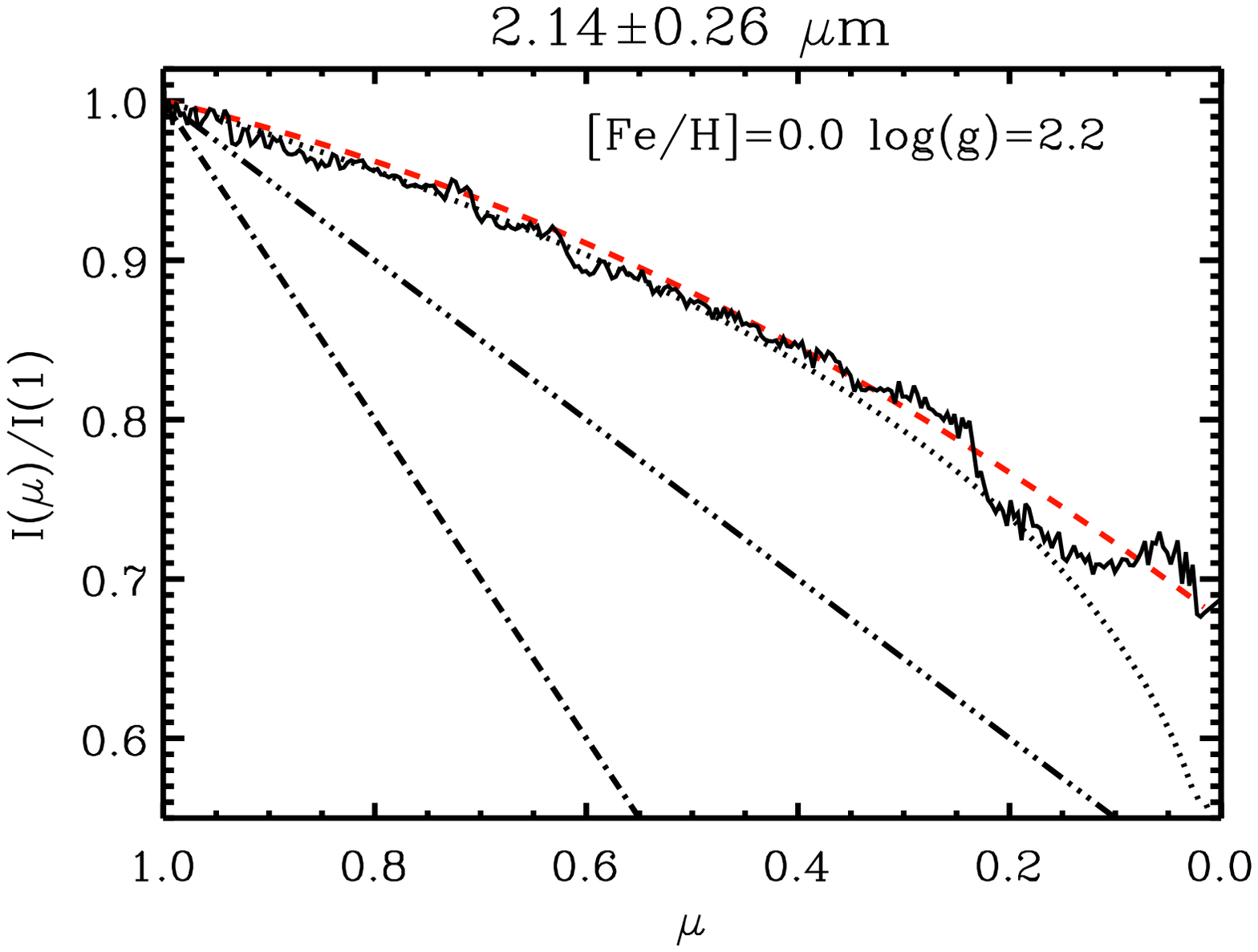} \\
     \includegraphics[width=0.5\hsize]{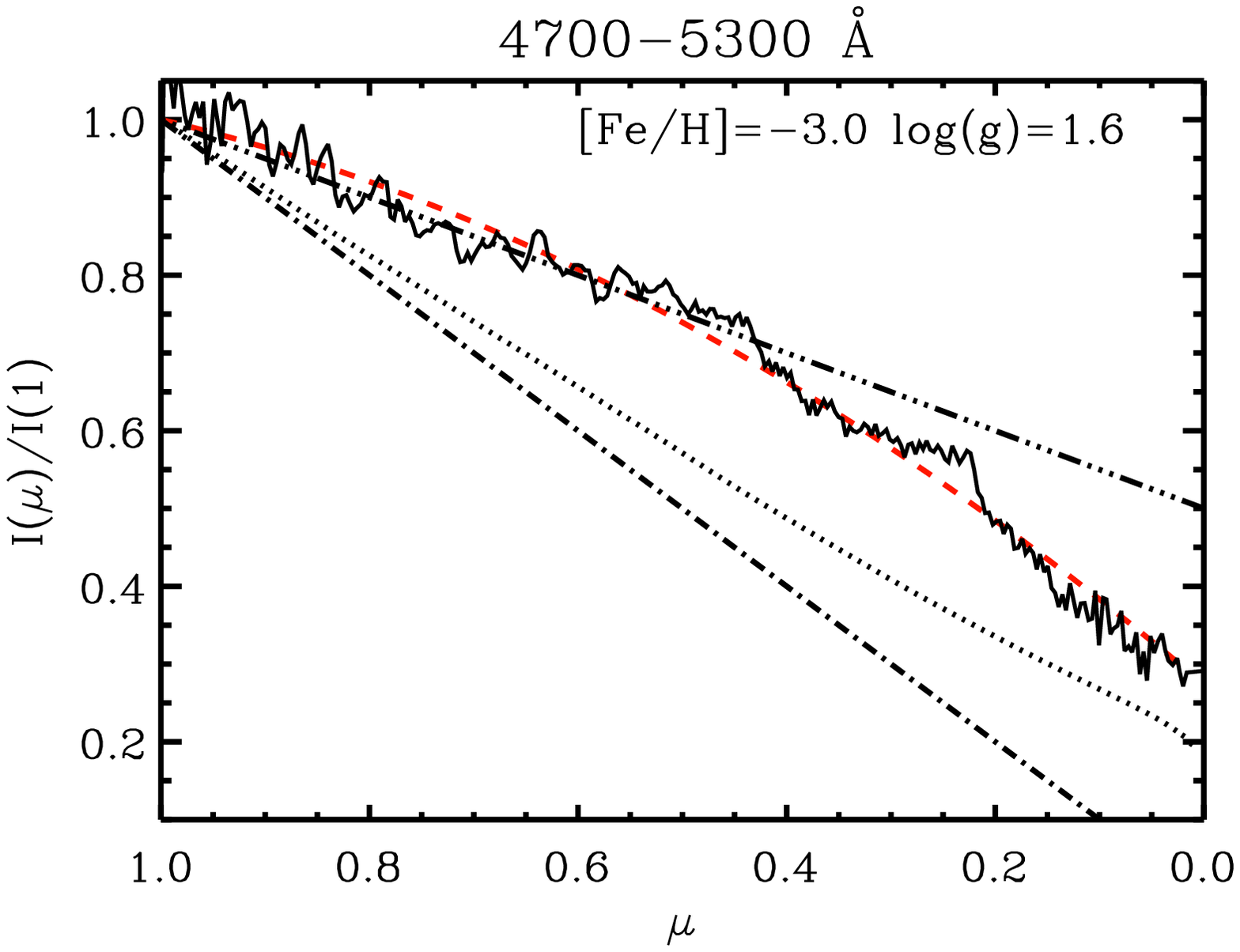} 
 \includegraphics[width=0.5\hsize]{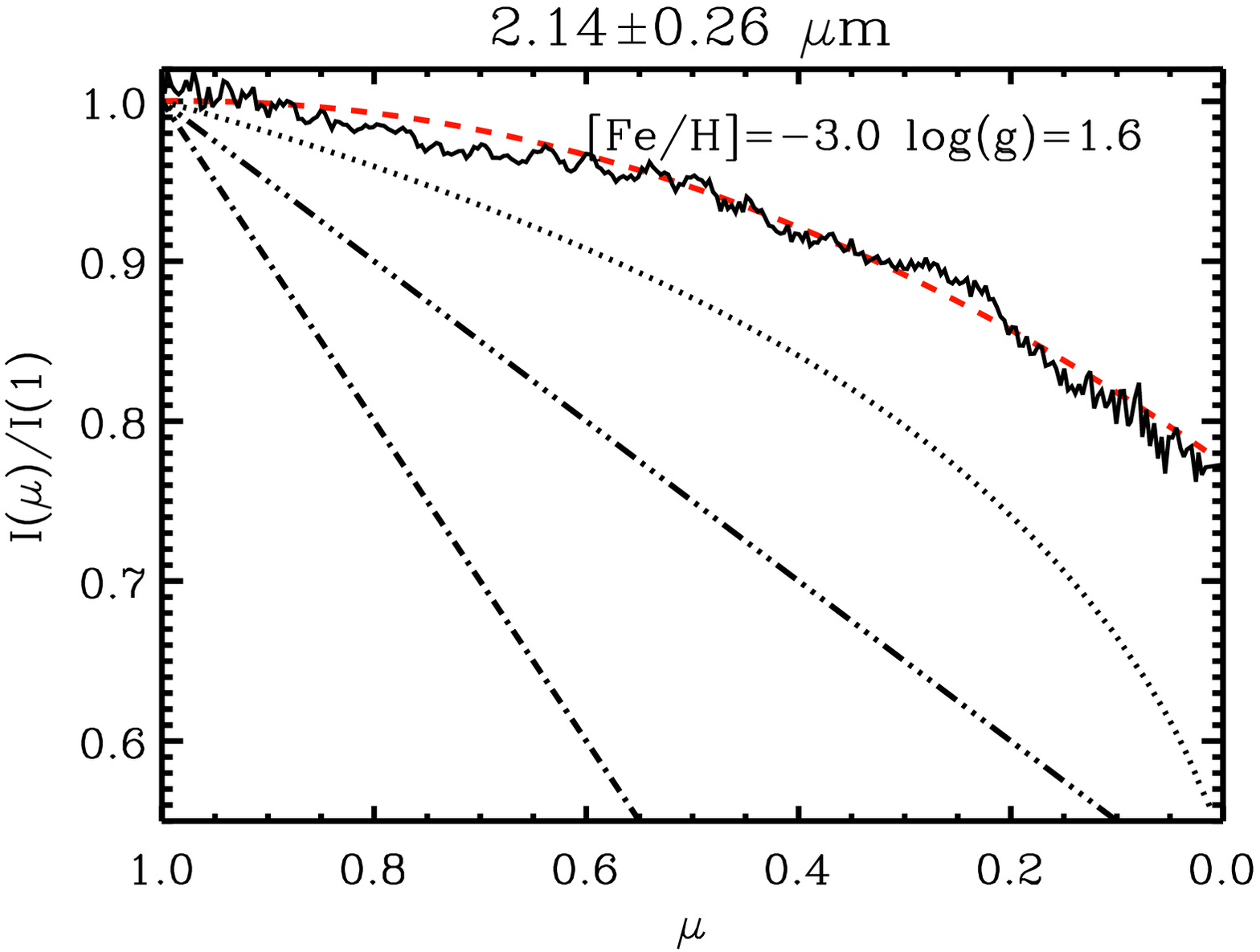} 
 \end{tabular}
     \caption{LD fits (red dashed line) obtained using the
       Eq.~\ref{claret_law} with $N=2$ for the RHD azimuthally average
       intensity profile (solid line) of two simulations of Table~\ref{simus}. The dotted line is the intensity profile computed with the 1D model having identical stellar parameters, input
data, and chemical compositions as the 3D simulation. A full LD (dash-dotted line), a
     partial LD (triple dot-dashed line) are also shown. 
          }
       \label{ld_fits}
  \end{figure*}

We have also computed classical 1D, LTE, plane-parallel,
hydrostatic {\sc MARCS} model atmospheres \citep{1975A&A....42..407G, 1997A&A...318..521A} with identical stellar parameters, input
data, and chemical compositions as the 3D simulations of
Table~\ref{simus}. With {\sc Optim3D}, we computed the emerging
intensities for the same regularly spaced 285~$\mu$-values used to get
the azimuthally averaged intensity profiles from the synthetic stellar disk images  and for the wavelength range spanned by both filters, assuming a micro-turbulent
broadening of 2~$\mathrm{km}\mathrm{s}^{-1}$. For each $\mu$-value, we integrated the monochromatic intensities over the filters'
ranges to obtain a 1D intensity
profile. Figure~\ref{ld_fits} illustrates the comparison between the average
intensity profile from a 3D simulation, together with its LD fit,
compared to a 1D model profile. The latter displays large
differences with respect to 3D profile. Moreover, full LD
($I(\mu)/I(1)=\mu$) and partial LD ($I(\mu)/I(1)=0.5\cdot\mu$) profiles largely diverge from both 1D and 3D profiles. The effects
on diameter determination from these visibility curves are analyzed in Section~\ref{Sectradii}.

\section{Interferometric predictions: departure from spherical symmetry} \label{SectVis}

We derive the interferometric observables to determine the impact of
the limb-darkening and granulation pattern on visibility curves and
closure phases, using the same method described in
\cite{2009A&A...506.1351C}. For each synthetic stellar disk image, we
calculated the discrete complex Fourier transform $z$. The visibility,
$vis$, is defined as the modulus $|z|$, of the Fourier transform
normalized by the value of the modulus at the origin of the frequency
plane, $|z_0|$, 
with the phase $\tan\varphi = \Im(z)/\Re(z)$, where $\Im(z)$ and
$\Re(z)$ are the imaginary and real parts of the complex number $z$,
respectively. We introduced also a theoretical spatial frequency scale
expressed in units of inverse solar radii (R$_\odot^{-1}$). The conversion between spatial frequencies expressed in the latter scale and in the more usual scale of $1/\arcsec$ is given by: 

\begin{equation}\label{eqvis1}
\nu~\left[\frac{1}{\arcsec}\right]=\nu~\left[\frac{1}{{\rm R}_\odot^{-1}}\right]\cdot d~[{\rm pc}]\cdot214.9
\end{equation}
where $\nu$ is the spatial frequency, 
214.9  is the astronomical unit expressed in solar radii, and
$d$ is the distance of the observed star. Also useful is the following
relation 
\begin{equation}\label{eqvis2}
\nu=\frac{B}{\lambda\cdot0.206265}
\end{equation}
where $\nu$ is the spatial frequency in arcsec$^{-1}$ at the observed
wavelength $\lambda$ in $\mu{\rm m}$ for the baseline $B$ of an
interferometer in meters. 

\subsection{Visibility curves}

\begin{figure*}
  \centering
   \begin{tabular}{c}
     \includegraphics[width=0.5\hsize]{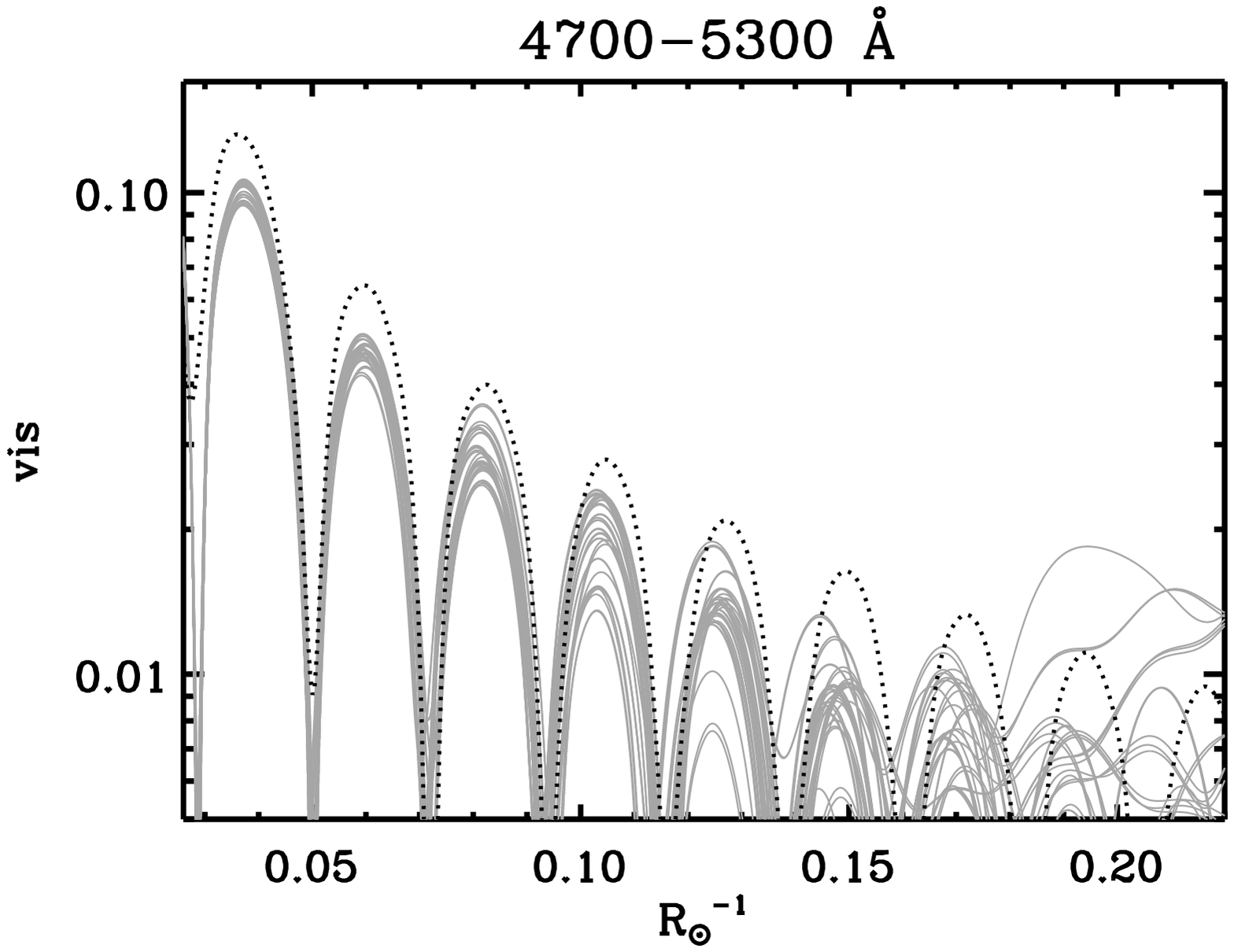} 
 \includegraphics[width=0.5\hsize]{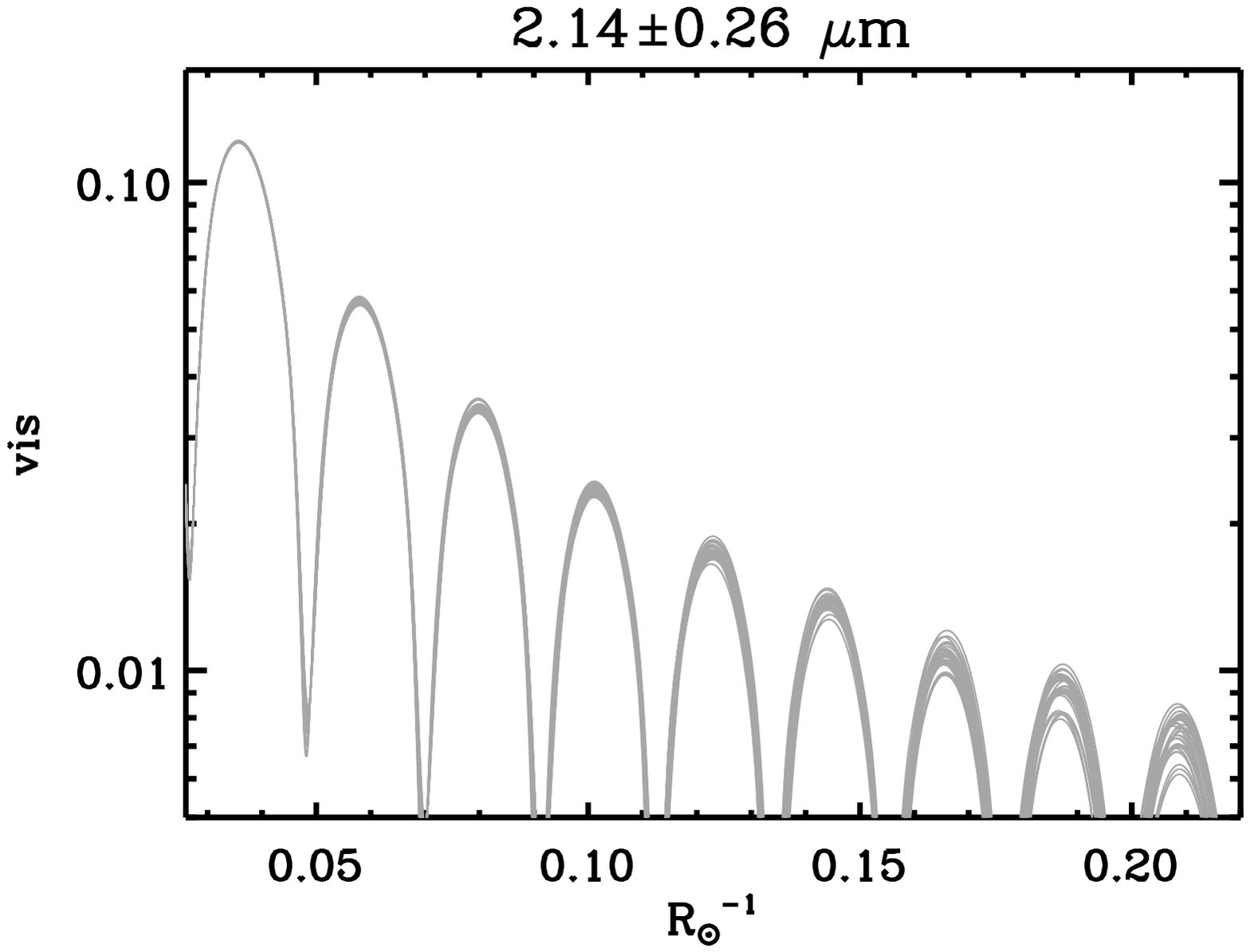} \\
  \includegraphics[width=0.5\hsize]{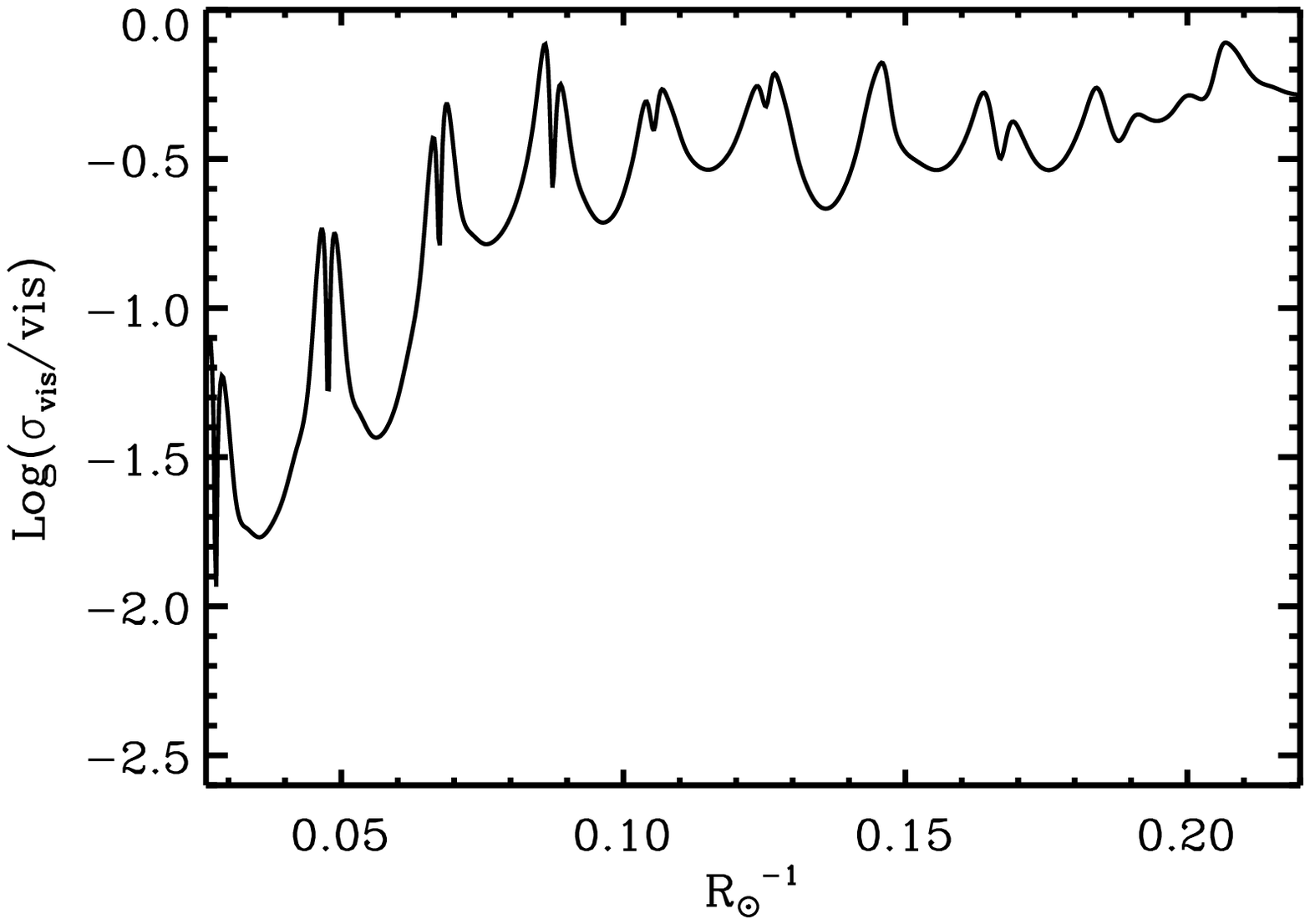} 
 \includegraphics[width=0.5\hsize]{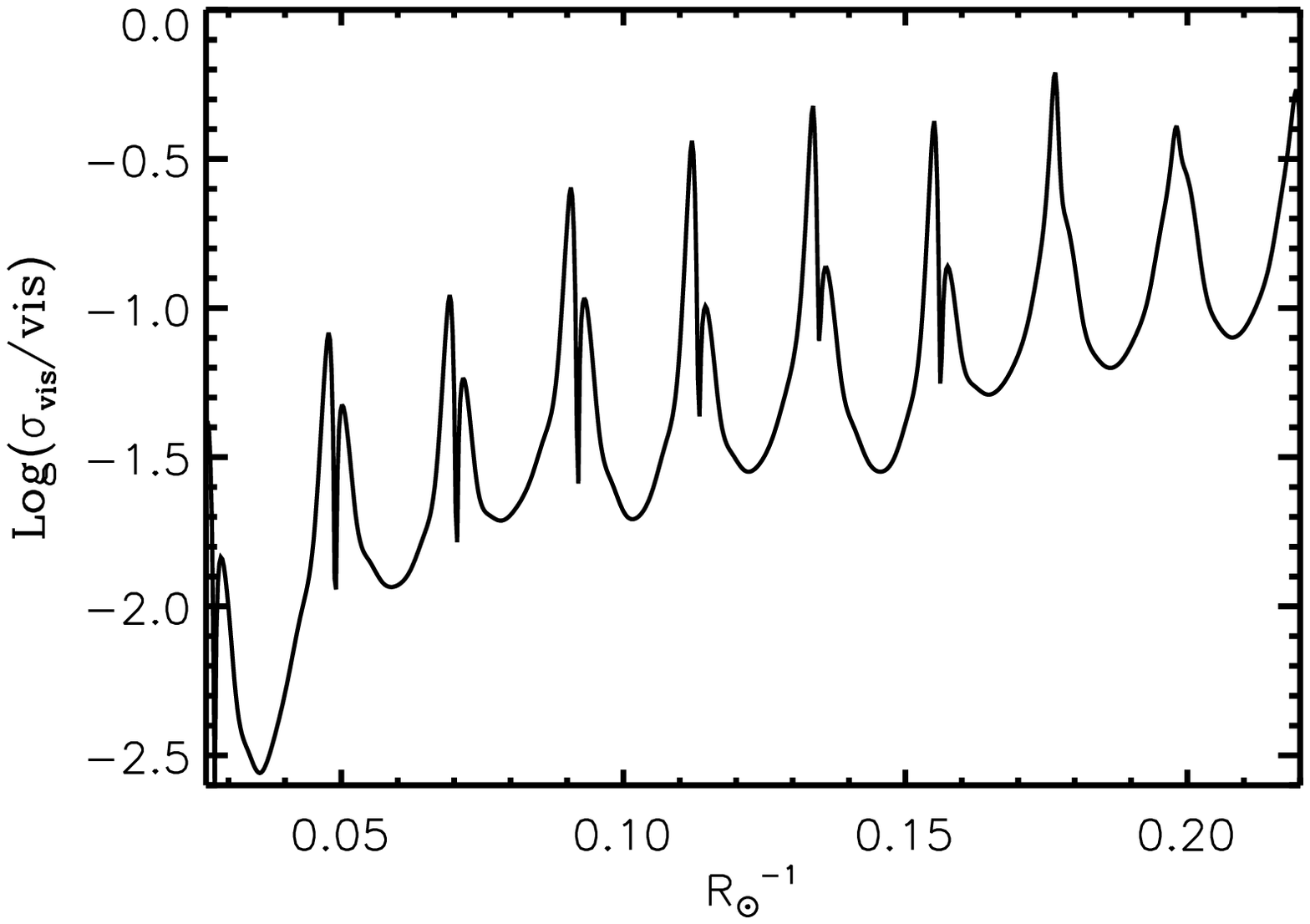} \\
 \end{tabular}
     \caption{\emph{Top row:} visibility curves from the RHD
       simulation with [Fe/H]=$-$3.0 and $\log g$=1.6 in the optical (left column) and
       infrared (right column) filters. The visibilities are computed
       for 36 different azimuth angles 5$^\circ$ apart (thin grey lines). The
       dotted line is an uniform disk scaled to match the same
       radius, i.e. the first null in visibility curve. The
       visibilities are displayed only longward of the first null
       visibility point (see text). A logarithm scale is used on y-axis. \emph{Bottom row:} visibility fluctuations with respect to the average value as a function of spatial frequencies.
          }
       \label{vis-fluct}
  \end{figure*}

The first null point of the visibility curve is mostly sensitive to
the radial extension of the observed object \citep[e.g.][and
 \citeauthor{2010A&A...515A..12C}, \citeyear{2010A&A...515A..12C} for
 an application to RHD simulations]{2001ARA&A..39..353Q}. In fact,
the radius of Table~\ref{simus} can be estimated using
R$_\star[\rm{R}_\odot]=\frac{\Theta[\rm{R}_\odot]}{2}=\frac{1.22}{2\cdot\nu[\rm{R}_\odot^{-1}]}$, where
$\nu[\rm{R}_\odot^{-1}]$ is the spatial frequency corresponding to the
first null point and $\Theta$ the angular diameter. This formula is related to the resolving power of a
telescope. Our predictions concerning the first lobe of the
visibility curves, however, cannot be used to fit the observed data because of
the geometrical limitations of our models. In our spherical tiling
model, based on \emph{box-in-a-star} simulations, the stellar radius
is effectively a free-parameter. Only 3D simulations in the \emph{star-in-a-box} configuration can be strictly used for this purpose
\citep{2009A&A...506.1351C}. On the other hand, the first null point and the second lobe of the visibility curves are sensitive to the limb-darkening \citep{1974MNRAS.167..475H}. These frequency points are then crucial to test RHD simulations against observations to recover the information on the temperature stratification in the external layers of the star. 

\begin{figure*}
  \centering
   \begin{tabular}{c}
     \includegraphics[width=0.4\hsize]{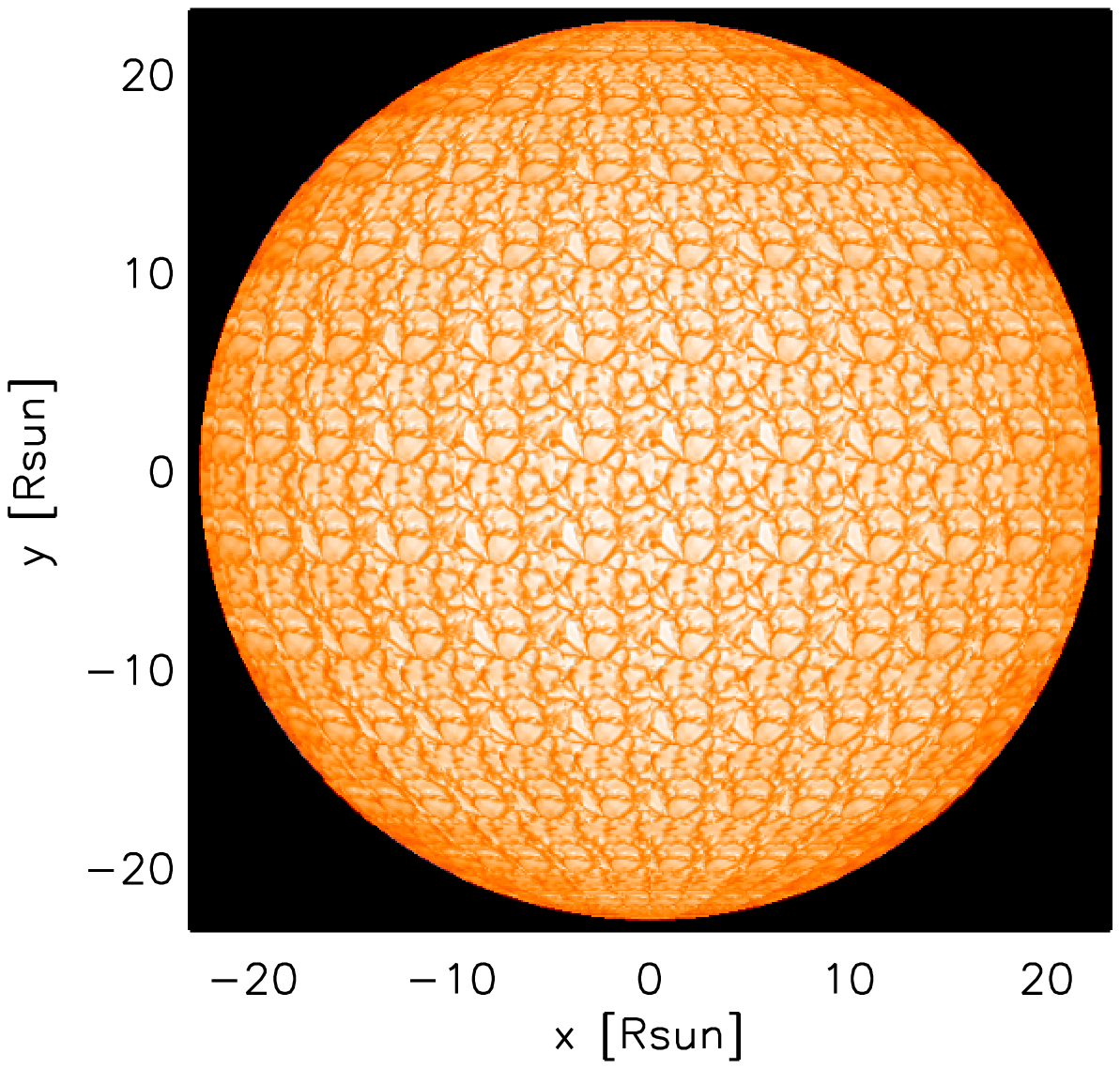} 
 \includegraphics[width=0.4\hsize]{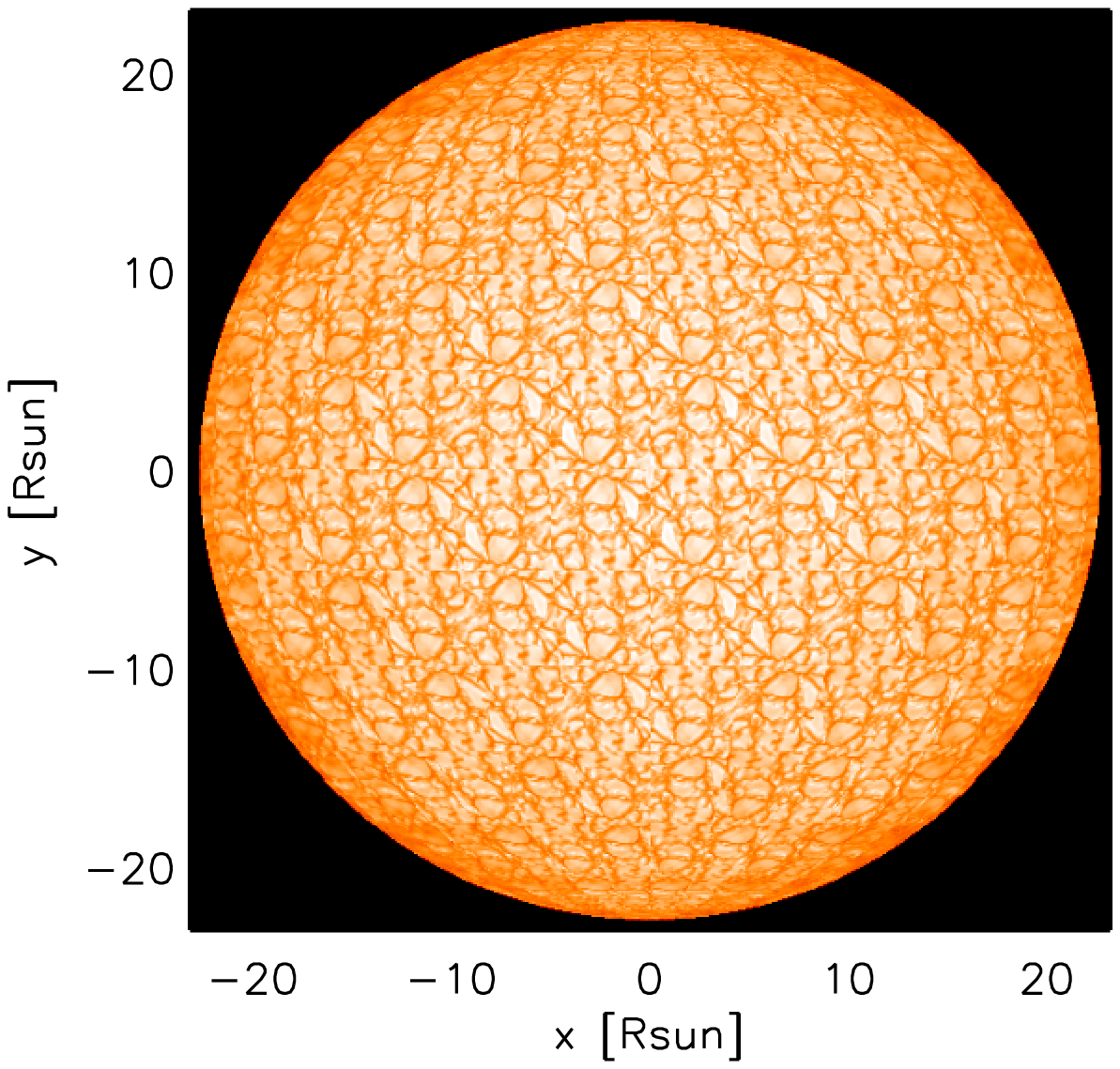} \\
  \includegraphics[width=0.5\hsize]{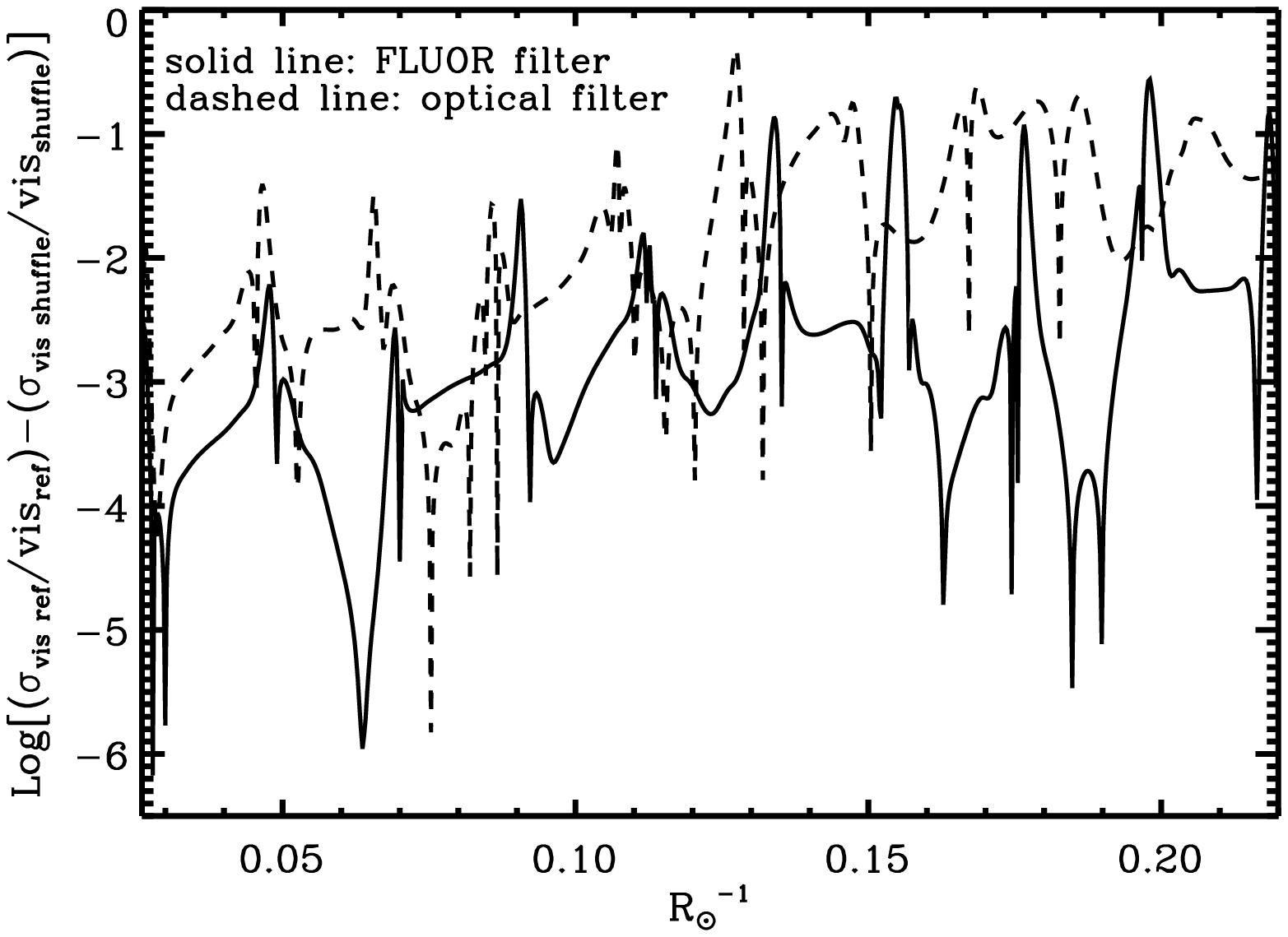}
 \end{tabular}
     \caption{\emph{Top left panel:} synthetic stellar disk image in the
       FLUOR filter for the simulation with [Fe/H]=$-$3.0 and $\log g$=1.6
       tiled using only one intensity map at different
       $\mu$-angles (see text for details). The granulation pattern is continuous. The
       intensity range is [5000,
         $0.11\times10^6$]\,erg\,cm$^{-2}$\,s$^{-1}$\,{\AA}$^{-1}$. \emph{Top
         right panel:} synthetic stellar disk image in the same filter and
       using the same intensity map but flipping the y-axis every
       second tile to remove the periodicity at the side of each box. \emph{Bottom panel:} Ratio of the fluctuations of the synthetic visibility curves derived from the above images: $\sigma_{\rm{ref}}/\rm{vis_{\rm{ref}}}$ refers to top left panel and $\sigma_{\rm{shuffle}}/\rm{vis_{\rm{shuffle}}}$ to top right panel.
          }
       \label{error}
  \end{figure*}

Figure~\ref{vis-fluct} (top panels) shows the visibility curves
computed for 36 different cuts through the centers of the stellar disk images
in Fig.~\ref{intensity_images}. This is equivalent to
generating different realizations of the stellar disk images with intensity maps computed for different sets of 
randomly selected snapshots. The comparison with the uniform disk
visibility in top left panel of Fig.~\ref{vis-fluct} shows that the
synthetic visibilities deviate greatly from circular symmetry and they
are systematically lower than the uniform disk model at the leftmost
spatial frequencies shown in the plot. This is due to a the limb darkening effect already visible
in Fig.~\ref{intensity_images}. Moreover, bottom panels
of Fig.~\ref{vis-fluct} shows the one $\sigma$ visibility
fluctuations, $F$,
with respect to the average value $\overline{{vis}}$
($F=\sigma/\overline{{vis}}$): the dispersion clearly increases with
spatial frequency. On the top of the second lobe, which is easy to
reach both in the optical and in the FLUOR filters with CHARA of a
typical red giants, the fluctuations are about 2$\%$ (optical) and 0.3$\%$ (FLUOR). In fact, the effect of line blanketing in FLUOR filter (Fig.~\ref{intensity_images}, right panel) is less important than the in the optical case and the image seems closer to a uniform disk.\\
In addition to this, the synthetic visibilities of the simulation at
$\log g=1.6$ and [Fe/H]=$-$3.0 show, in the optical filter, the
largest deviation from UD in the range $0.17-0.22$~
R$^{-1}_\odot$. This is due to the small scale structure on the model
stellar disk shown in \cite{2010A&A...515A..12C}. This frequency range
corresponds to surface structures of $2.27-2.94$ $\rm{R}_\odot$ or
$1500-2200$ Mm, approximatively the size of one of the granules as shown in
Fig.~\ref{limb}. 

Observing red giant stars at high spatial resolution is thus crucial
for two reasons, at least:

\begin{itemize}
\item[(1)] To further test the limb darkening predicted by our RHD simulations. For this purpose, it would be worthy searching for angular visibility variations, observing with the same telescope configuration 
(covering spatial frequencies higher than the first lobe) and using the Earth rotation in
 order to span multiple position angles. Assuming  measurement errors
 below 2$\%$, for visibilities of $\sim10\%$, one night would be
 enough. For comparison, the instrumental error of VEGA on CHARA is $\sim$1$\%$ \citep{2009A&A...508.1073M}.
\item[(2)] To characterize the granulation pattern. This requires
 observations at very high spatial frequencies that may be possible
 only in the optical range. In fact, considering that the maximum
 baseline of CHARA is 331 meters \citep{2005ApJ...628..453T},
 for a red giant with an apparent radius of 2.7 mas
 \cite[e.g. HD~214868][]{2010ApJ...710.1365B}, a baseline of
 $\sim$270 (1200) meters is necessary to probe the seventh lobe at
 0.5 (2.14) $\mu$m.
\end{itemize}

Our method of constructing realizations of red giant stellar disk images (Section~\ref{Sectionortho})
inevitably introduces some discontinuities between neighbouring tiles by randomly selecting temporal snapshots and by cropping intensity maps at high latitudes and longitudes.
We checked whether these artifacts produce significant fluctuations in the theoretical visibility curves.
These discontinuities in the granulation pattern stem from using random snapshots of the simulation sequences. We
explored the impact of such discontinuities at the edges of neighbouring tiles by
constructing two mock stellar disk images (Fig.~\ref{error}). 
To generate the first image, we tiled a sphere with intensity maps computed for one single simulation snapshot: in this case, the continuity of the granulation pattern is ensured by the
the periodic horizontal boundary conditions of the simulation.
For the second image, we used the same intensity map as for the previous case but flipping the y-axis every second tile: in this way, we generated a granulation pattern that is statistically similar to the one in the first image in terms of fine spatial structures but at the same time has discontinuities at the boundaries between tiles (top right panel). 
We then computed the visibility curves for 36 different angles 5$^\circ$ apart and derived the visibility fluctuations
(Fig.~\ref{error}, bottom panel). Both in the optical and in the FLUOR
filter, the visibility fluctuations caused by the disentangled
periodicity at the sides of each boxes have a contribution about two
orders of magnitude smaller than the signal observed in Fig.~\ref{vis-fluct}. We conclude that our approach to construct the orthographic projection with random snapshots have a smaller effect on the visibility curves than the signal originated by the inhomogeneities of the stellar surface.

\begin{figure}
  \centering
     \includegraphics[width=1.0\hsize]{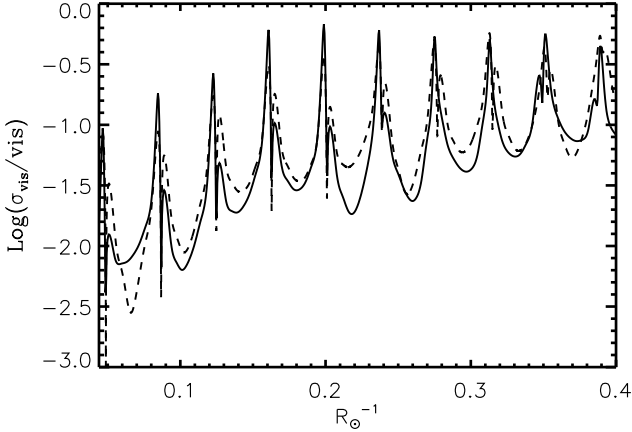} 
     \caption{Visibility fluctuations with respect to the average value as a function of spatial frequencies for the RHD
       simulation with [Fe/H]=0.0 and $\log g$=2.2 in the FLUOR filter. The dashed line corresponds to fluctuations of visibility computed taking into account the bandwidth smearing effect (see text and Eq.~\ref{eq_smearing}) while the solid line to fluctuations of visibility computed without bandwidth smearing.
          }
       \label{smearing}
  \end{figure}

In relatively large filters, such as FLUOR, several spatial frequencies are simultaneously
observed by the interferometer. This effect is called bandwidth smearing. \cite{2003A&A...408..681K, 2003A&A...404.1087K} show that this effect is negligible for squared visibilities larger than 40$\%$ but it is important spatial frequencies close to the first
minimum of the visibility function. To account for this effect, \cite{2004A&A...413..711W} proposed to compute averaged squared visibility amplitudes, $\langle vis^2\rangle$, as

\begin{equation}\label{eq_smearing}
\langle vis^2\rangle=\frac{\int_{\lambda_0}^{\lambda_1} vis^2_{\lambda}d\lambda}{\int_{\lambda_0}^{\lambda_1} T^2_\lambda F^2_\lambda d\lambda}
\end{equation}

where $vis^2_{\lambda}$ is the squared visibility at wavelength $\lambda$, $\lambda_0$ and $\lambda_1$ are the filter wavelength limits, $T_\lambda$ is the transmission curve of the filter and $F_\lambda$ is the flux at wavelength $\lambda$ (see Fig.~\ref{filters}). \\
In our case, first we generated a synthetic stellar disk image at every wavelength and then we computed the visibility curves $vis^2_{\lambda}$ for 36 different cuts through the centers of the stellar disk images. Then we applied Eq.~\ref{eq_smearing} to obtain the averaged squared visibilities and finally we derived the visibility fluctuations with respect to the average values. Figure~\ref{smearing} shows that there is not a clear difference between the visibility fluctuations computed with and without taking into account the bandwidth smearing effect.

\begin{figure*}
  \centering
   \begin{tabular}{cc}
     \includegraphics[width=0.35\hsize]{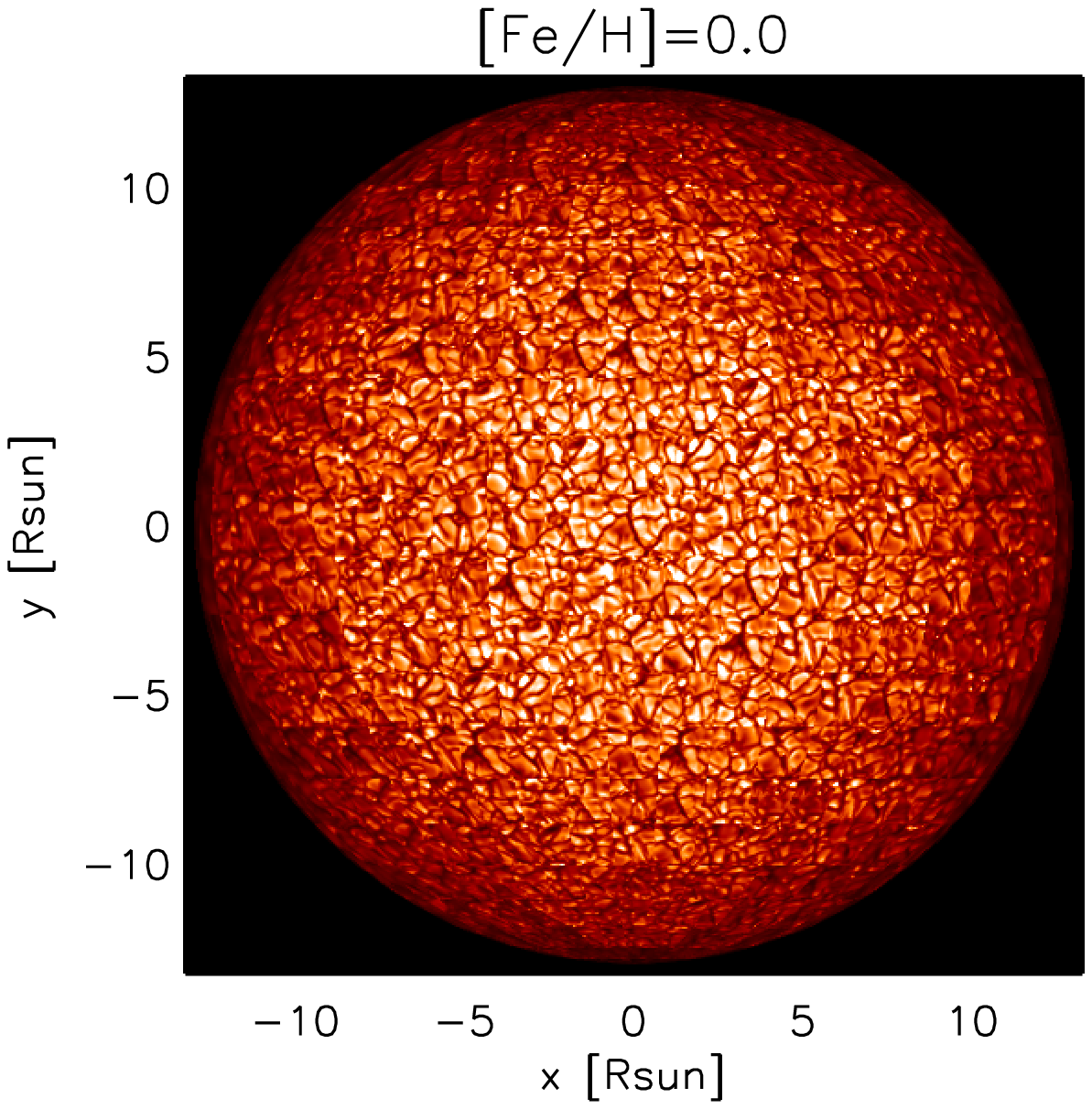} 
 \includegraphics[width=0.35\hsize]{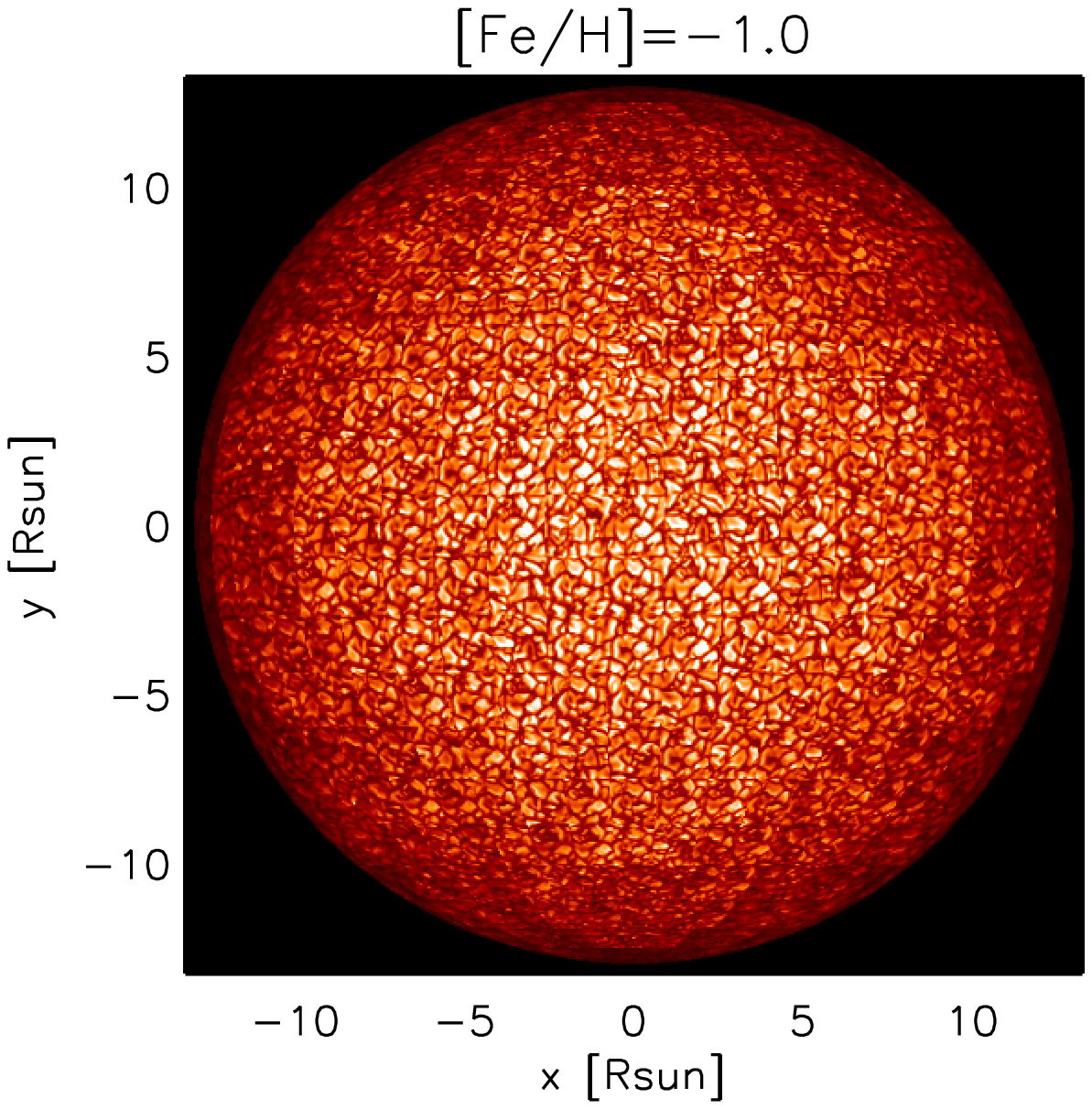} \\
    \includegraphics[width=0.35\hsize]{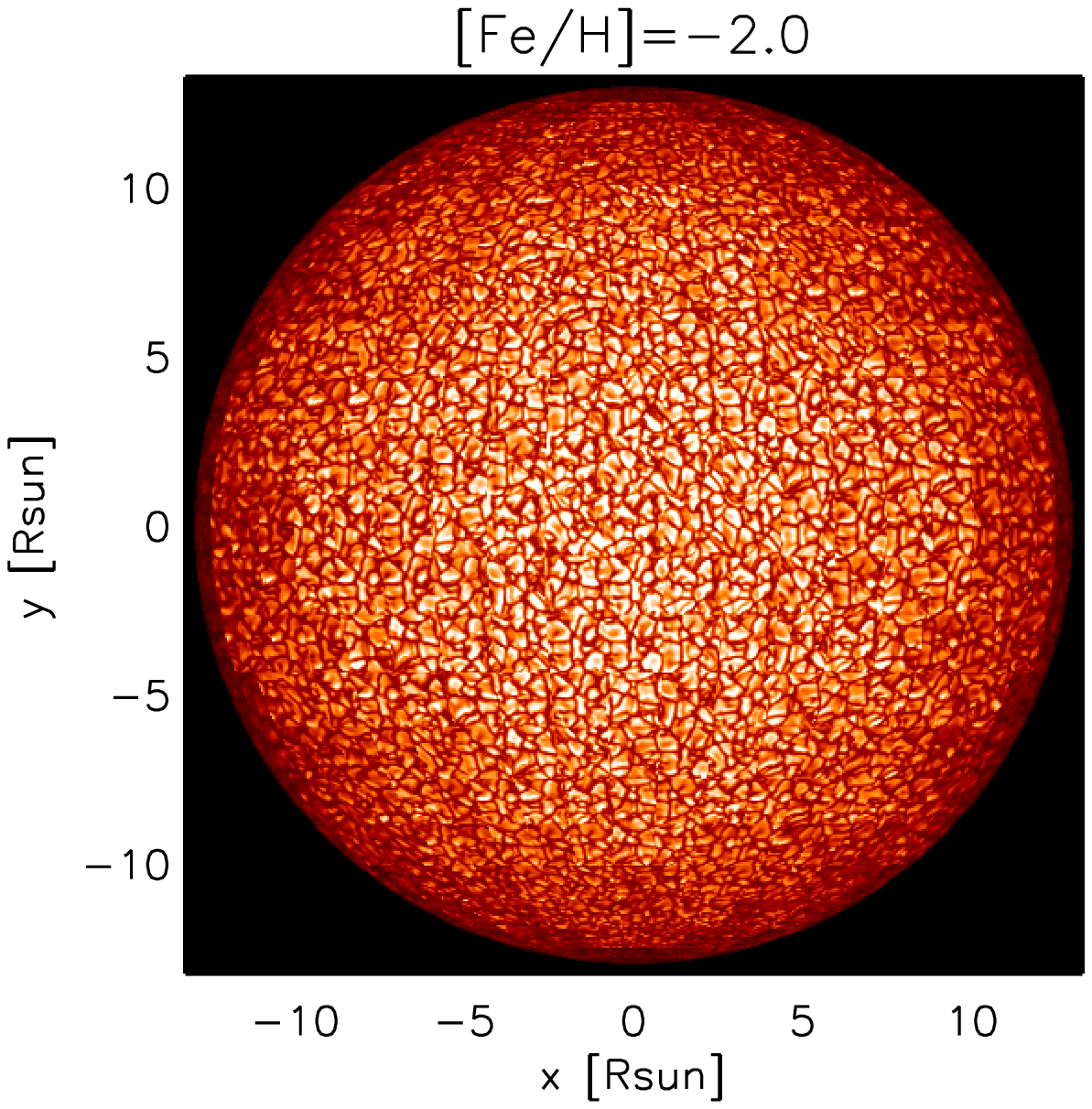} 
 \includegraphics[width=0.35\hsize]{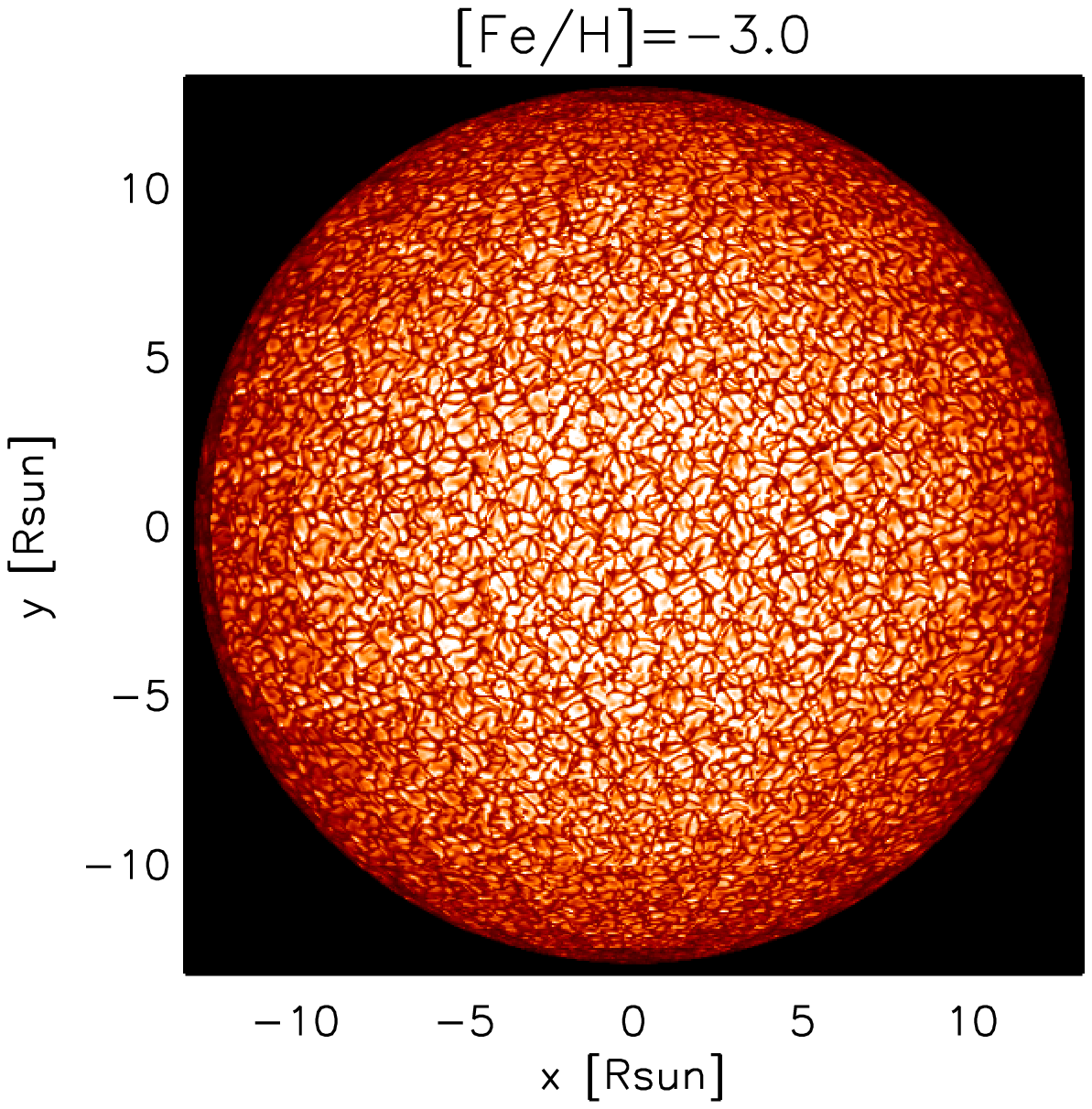} \\
      \includegraphics[width=0.5\hsize]{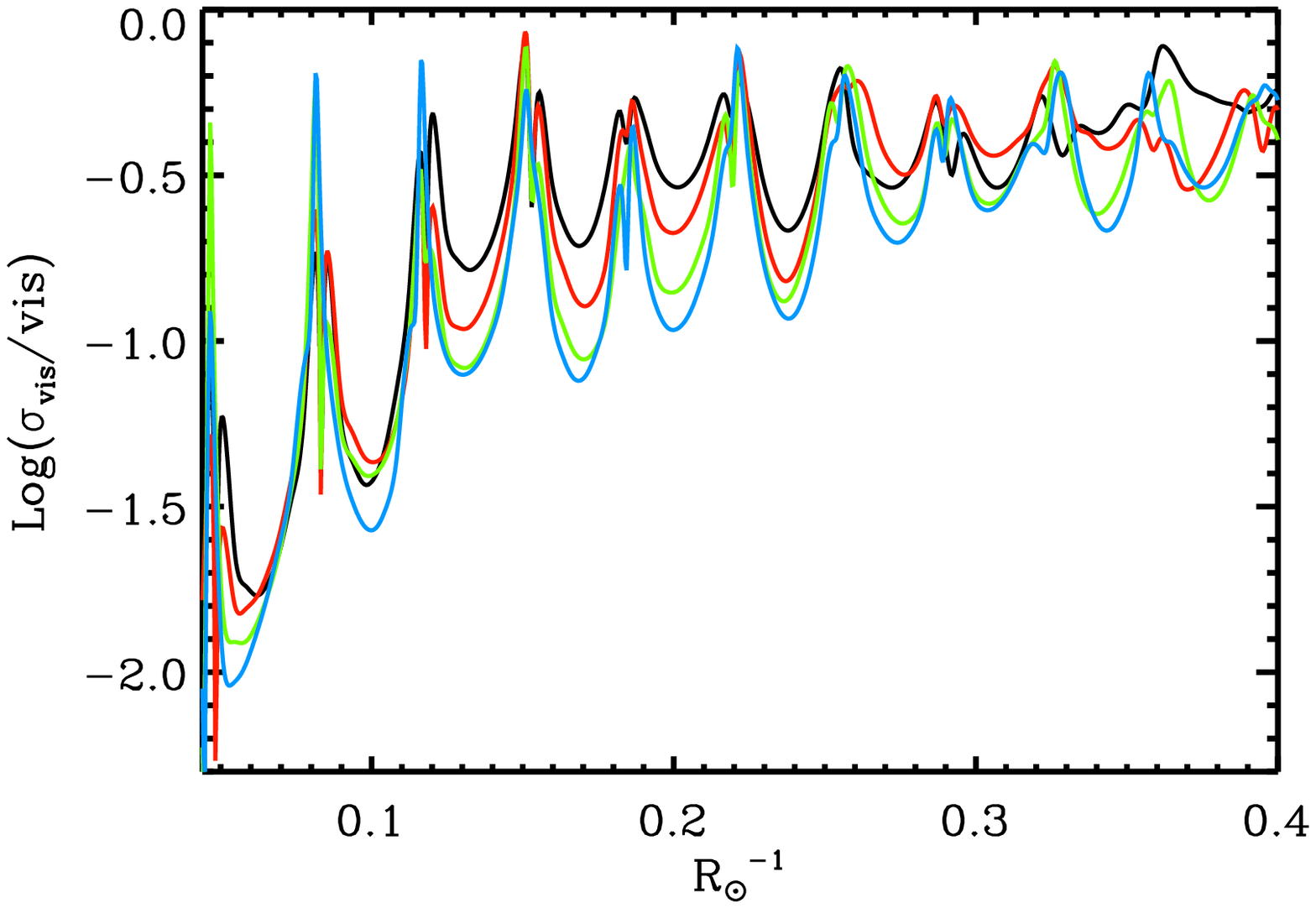} 
 \end{tabular}
     \caption{\emph{Top and central rows:} synthetic stellar disk
       images in the optical filter for the four simulations of
       Table~\ref{simus} having ${\rm log}~g$=2.2. The intensity range is [5000, $1.5\times10^6$]\,erg\,cm$^{-2}$\,s$^{-1}$\,{\AA}$^{-1}$ for [Fe/H] = 0.0 and -1.0; [5000, $2.0\times10^6$]\,erg\,cm$^{-2}$\,s$^{-1}$\,{\AA}$^{-1}$ for [Fe/H] = -2.0 and -3.0. \emph{Bottom panel:} fluctuations of the synthetic visibility curves derived from the above images. The colors have the same meaning as in Fig.~\ref{filters}. We dropped the first lobe of the visibility curves as in Fig.~\ref{vis-fluct}.
          }
       \label{intensity_metallicity}
  \end{figure*}

The simulations of \cite{2007A&A...469..687C} showed that the
predicted characteristic size of granules decreases with the
metallicity of the star. Therefore, we focus now on the
simulations of Table~\ref{simus} with the same surface gravity, ${\rm
 log}~g$=2.2, and different [Fe/H] and effective temperature. We
considered only the optical filter that show larger (and thus easier
to detect) visibility fluctuations. Figure~\ref{intensity_metallicity}
displays the synthetic stellar disk images and the visibility fluctuations
derived from the visibility curves. The solar metallicity simulation
has almost systematically smaller fluctuations (except for the 3rd
lobe at $\rm{R}^{-1}_\odot\sim0.10$) at all frequencies until
$\rm{R}^{-1}_\odot\sim0.26$ (i.e. the 6th lobe, that is close to the
baseline extension limit of CHARA interferometer for a 2.7 mas
star). The difference between [Fe/H]=0.0 (black curve) and [Fe/H]=$-$3.0
(light blue curve) is about 1$\%$ (same order of the instrumental
error of VEGA) on the top of the second lobe
($\rm{R}^{-1}_\odot\sim0.06$) with peaks of about $1.2$--$1.7$~$\%$ at
$\rm{R}^{-1}_\odot\sim0.17$ and $0.20$, respectively. We emphasize that it
is possible to study the metallicity dependence on visibility curves
to characterize the granulation pattern of red giant stars. It is however important to carefully consider specific instrumental effects that could affect the visibility fluctuations. This can
be already done with present day interferometers in addition to image
reconstruction \citep{2010Berger,2010arXiv1007.4473M} which eventually
will be
able to measure the size of granules directly.

\subsection{Estimation of diameter correction: 1D versus 3D models}\label{Sectradii}

Interferometric data are often fitted with a simple uniform disk model
that, although unphysical, has the advantage of immediately telling if
a star is resolved, after which limb-darkening corrections are applied.
Thus, to estimate the angular diameter from
measured visibilities it is necessary to know the intensity
distribution of the stellar disk and, for this purpose, parametric
(e.g. full or partial limb darkened) or photospheric one
dimensional models (e.g. MARCS) models are used. 

We used the intensity profiles obtained from 3D and 1D models (see 
Section~\ref{Sectlimb} and Fig.~\ref{ld_fits}) to estimate the
correction on the uniform disk diameter determination. We derived the visibility 
law $V_{\lambda}\left(B,\Theta\right)$ from the intensity profile 
$I\left(\lambda,\mu\right)$ using the Hankel integral:
\begin{equation}\label{hankel}
V_{\lambda}\left(B,\Theta\right)=\frac{1}{A}\int^1_0I\left(\lambda,\mu\right)J_0\left(\frac{\pi B\Theta}{\lambda}\sqrt{1-\mu^2}\right)\mu d\mu 
\end{equation}
where $\lambda$ is the wavelength in meters (in this case the central 
wavelength of the filters in Fig.~\ref{filters}), $B$ is the baseline in 
meters, $\Theta$ is an arbitrary angular diameter in radians (we assume 2 
mas), $J_0$ the zeroth order of the Bessel function, $\mu=cos(\theta)$ as 
defined in Section~\ref{Sectlimb}, and A the normalization factor:
\begin{equation}
A=\int^1_0I\left(\lambda,\mu\right)\mu d\mu
\end{equation}

\begin{figure*}
  \centering
   \begin{tabular}{cc}
     \includegraphics[width=0.5\hsize]{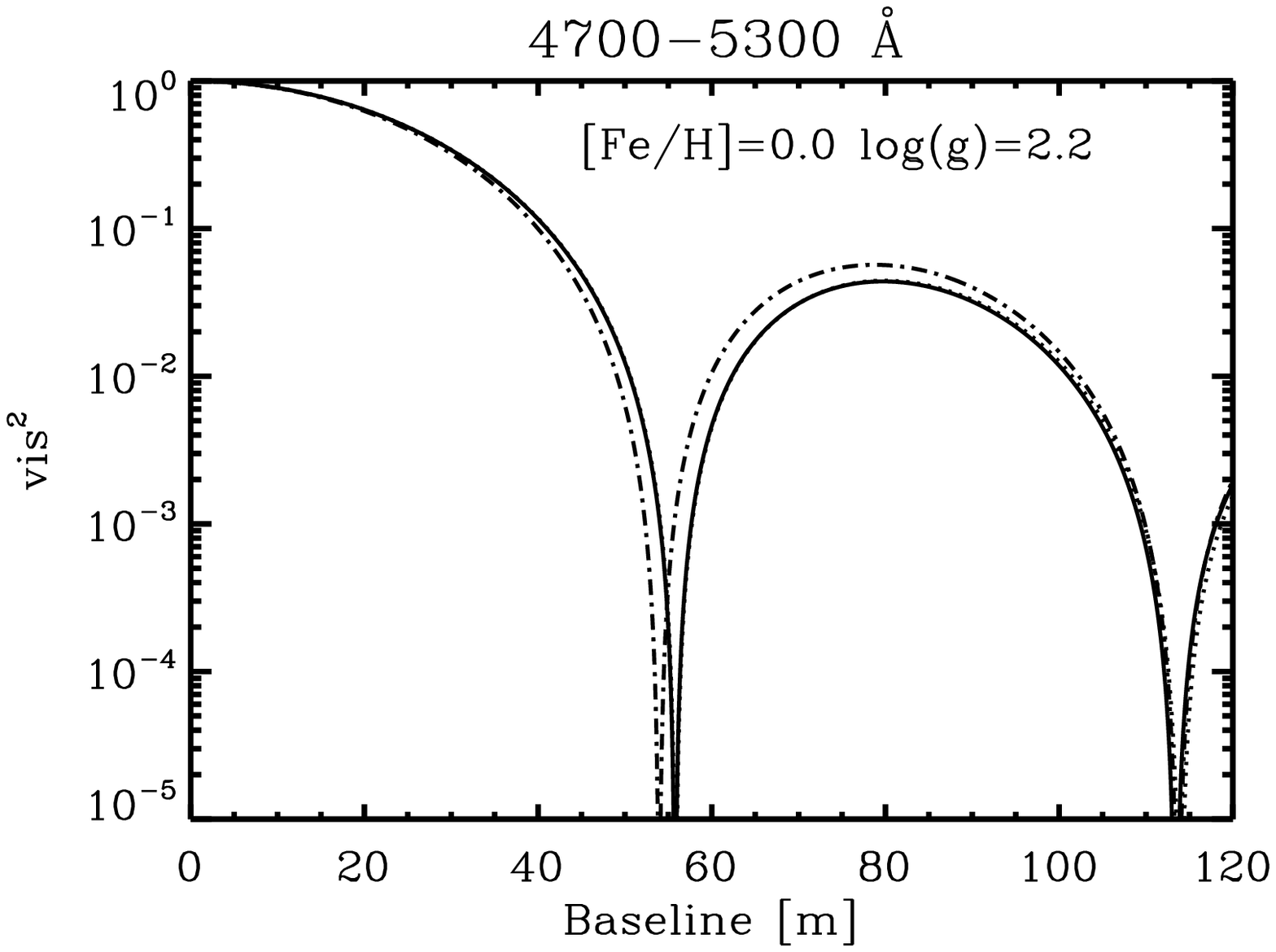} 
 \includegraphics[width=0.5\hsize]{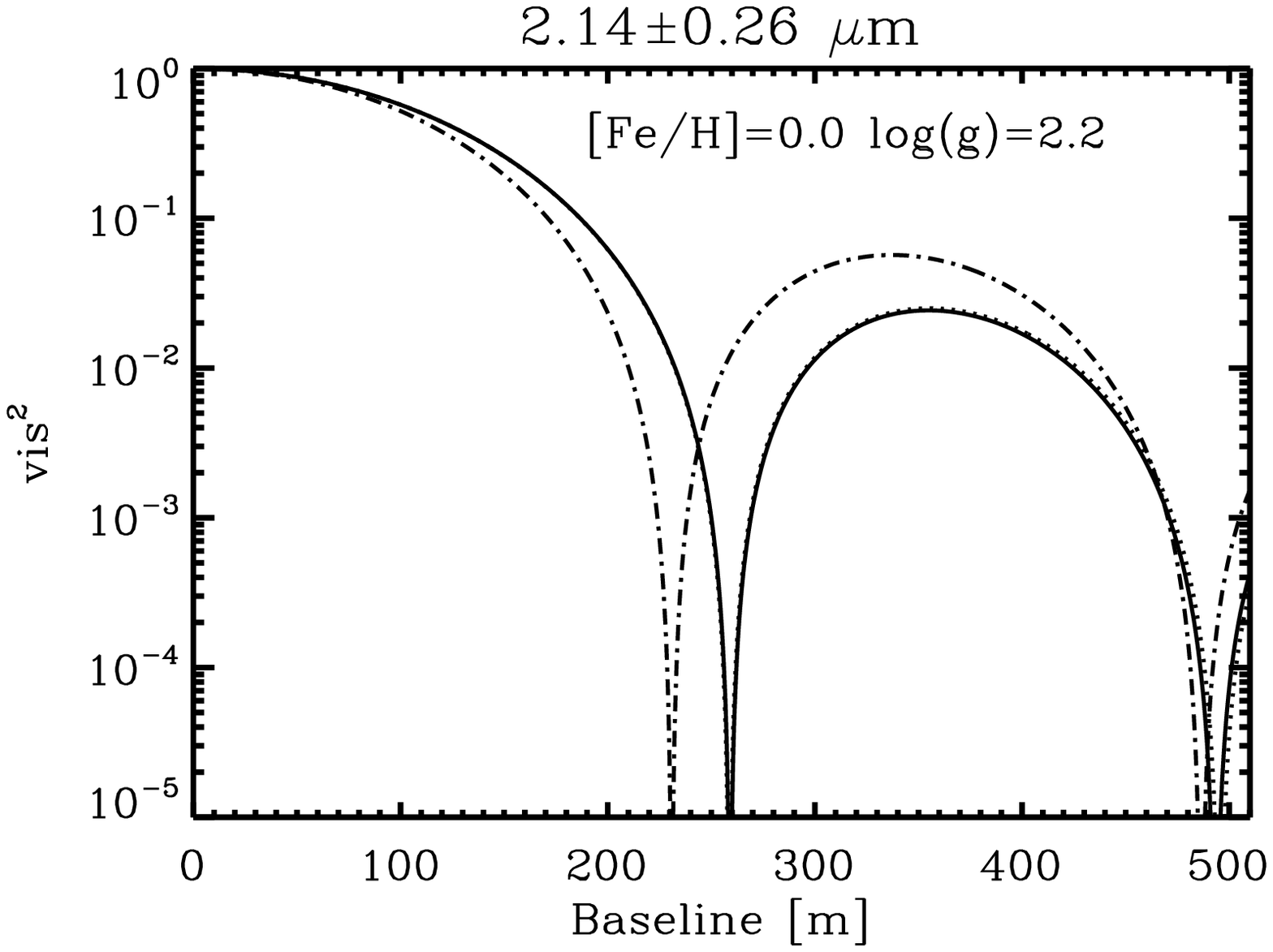} \\
    \includegraphics[width=0.5\hsize]{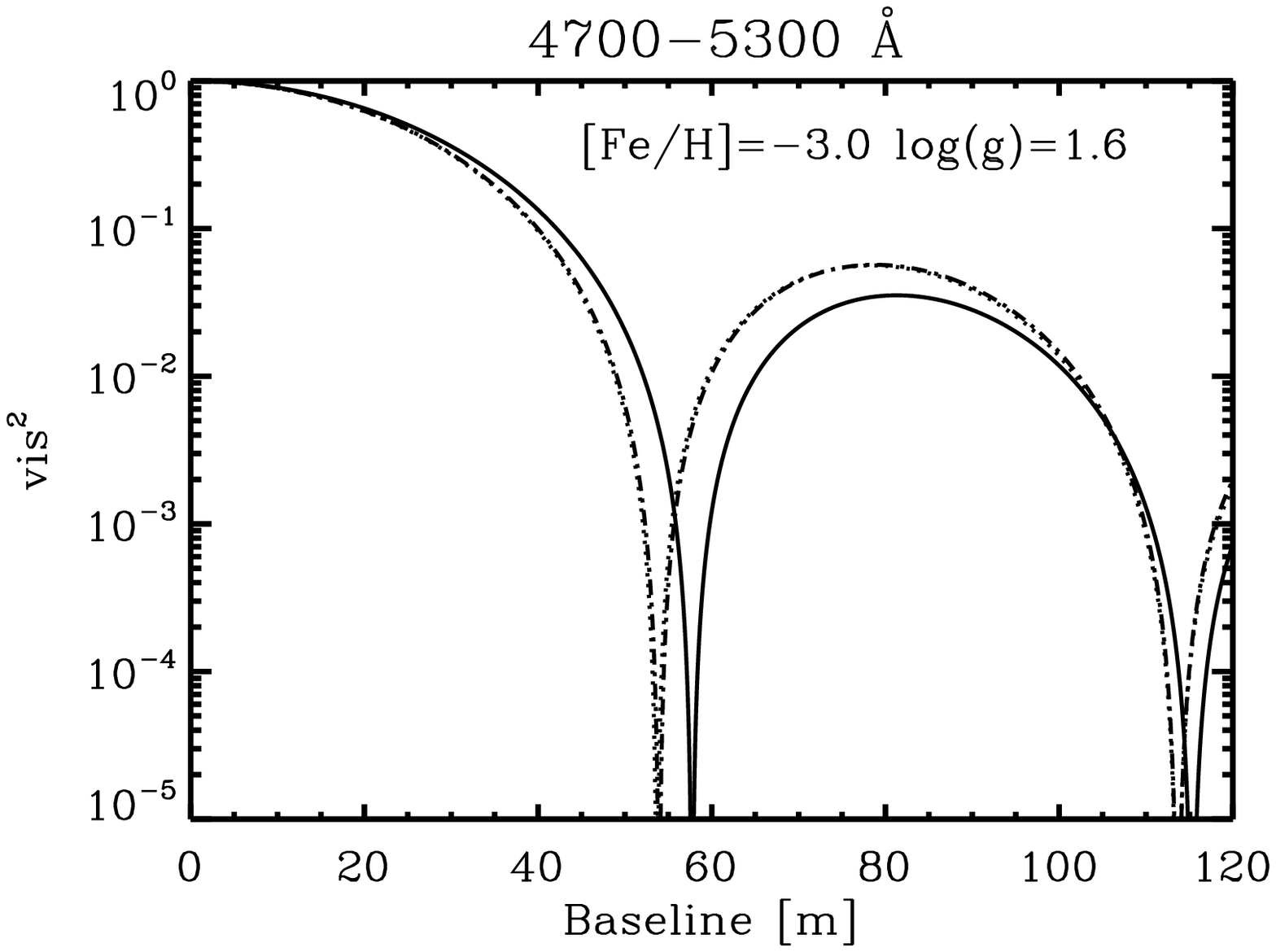} 
 \includegraphics[width=0.5\hsize]{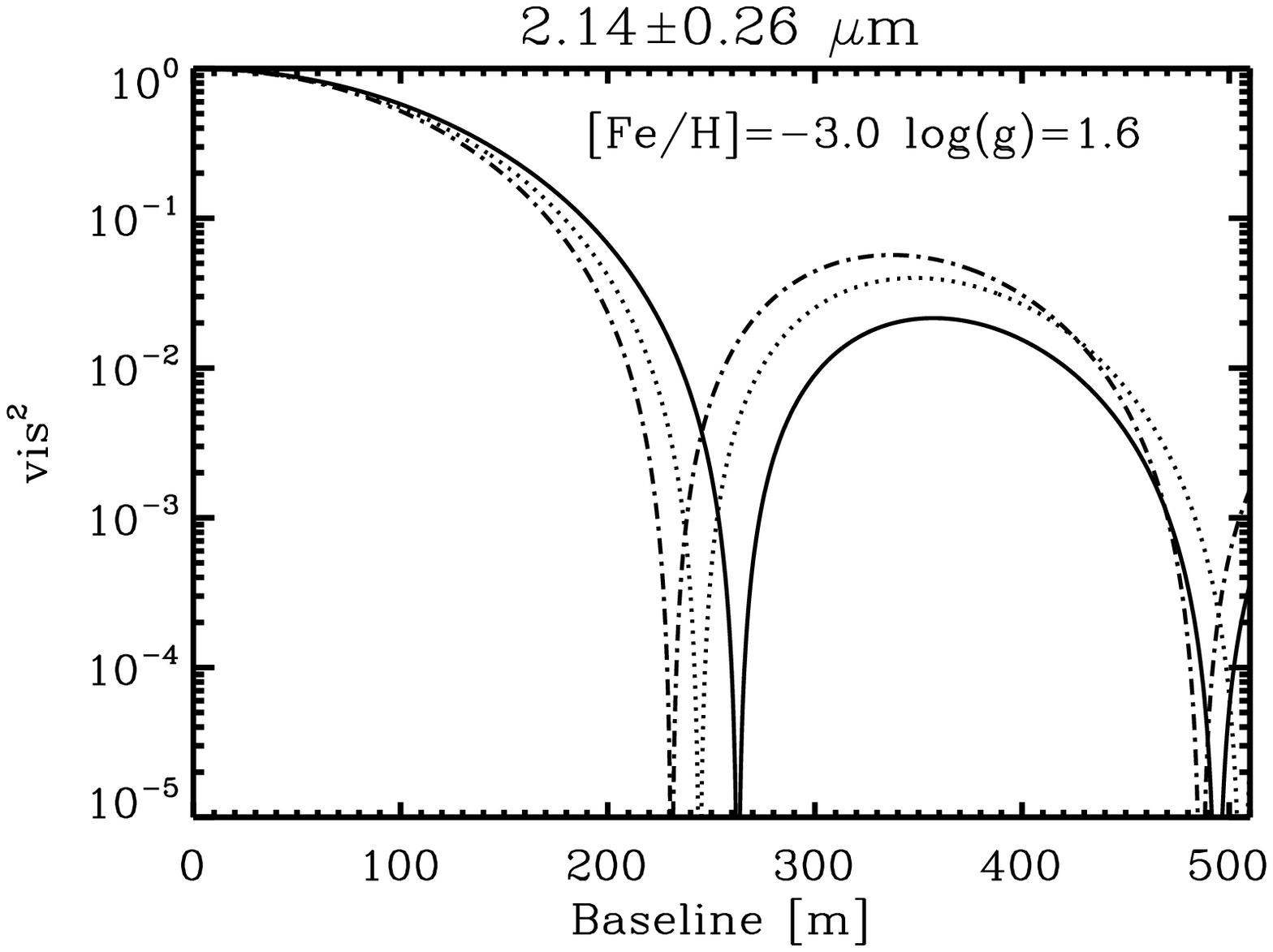} 
 \end{tabular}
     \caption{Visibility curves computed with Eq.~\ref{hankel} for
       and angular diameter of 2 mas and for the 3D (solid line), 1D
       (dotted line), and fully limb darkening (dash-dotted line)
       intensity profiles of Fig.~\ref{ld_fits}. In top panels, the
       dotted and solid lines are almost overlying, while in bottom
       left panel the dotted line is almost overlying the dash-dotted
       one. A logarithm scale is used on y-axis. Synthetic visibilities in these plots are not realistic near the nulls but are intended only for model-to-model comparison.
     }
       \label{vis_3D-1D}
  \end{figure*}

Figure~\ref{vis_3D-1D} shows the visibilities computed from the intensity 
profiles of Fig.~\ref{ld_fits}. For typical red giant stars having $4600 \lesssim \teff \lesssim
5100$~K (compare with Table~\ref{simus}) differences in angular diameters 
vary from $\sim -3.5\%$ to $\sim 1\%$ in the optical, and are roughly in the 
range between
$-0.5$ and $-1.5\%$ in the infrared (Table~\ref{table2}), the corresponding 
change in effective temperature being $\Delta \teff / \teff = 1 - \sqrt{ \Theta_{\rm{3D}}/ \Theta_{\rm{1D}}}$. These corrections depend on wavelength, the deviations being 
moderately stronger in the optical than in the infrared. Differences 
between the 1D and the mean 3D atmospheric structures and inhomogeneities of 
red giant stars have been already studied by means of 3D hydrodynamical 
simulations 
\citep{2007A&A...469..687C}. These differences have a significant effects on
the predicted strengths of spectral lines as well as on the center-to-limb 
profiles. As a consequence, when using LD laws obtained from 1D models 
\citep[e.g.][]{2000A&A...363.1081C} to determine angular diameters from 
observed data, one must take into account that 3D models usually predict a 
lower center-to-limb variations with respect to 1D, thus implying smaller 
radii and moderately hotter effective temperatures.

While the impact of the 
corrections in Table~\ref{table2} is usually not dramatic, they are not negligible to properly
set the zero point of the effective temperature scale derived by mean of this 
fundamental method.  In particular, when future long baseline interferometers 
working in the optical will be able to (partly) resolve (very) metal-poor 
stars, it will be fundamental to take those corrections into account for a 
correct derivation of their diameters. Moreover, the model dependence on the diameter
determination has also an impact on the reliability of the existing
catalogues of calibrator stars for interferometry that are based on
red giants with similar parameters to those studied in this
work. In fact, the formal
high accuracy reached on diameter determination \citep[a fraction of
 1$\%$,][]{2010Merand} is based only on 1D models, while we show that
the correction reported in Table~\ref{table2} can have a sizeable impact.

Our results in Table~\ref{table2} are in qualitative agreement with those obtained by 
\cite{2002ApJ...567..544A}, \cite{2005ApJ...633..424A} and 
\cite{2006A&A...446..635B} for dwarf stars, though the lower surface gravity 
of red giants might also play a role, since the size of surface related 
convective structures is roughly inversely proportional to the gravity. 
As we already discussed, the size of these inhomogeneities have an impact on 
the average intensity profile and thus on the interferometric observables.

\begin{table}
\begin{minipage}[t]{\columnwidth}
\caption{Ratio between the limb-darkened diameters recovered using 
 1D MARCS models ($\Theta_{\rm{1D}}$) or 3D simulations
 ($\Theta_{\rm{3D}}$) and the corresponding 
change in effective temperature $\Delta \teff$ for the RHD simulations of Table~\ref{simus}.}
\label{table2}
\centering
\renewcommand{\footnoterule}{}  
\begin{tabular}{c|cc|c|c}
\hline \hline
$\lambda$ [$\mu$m] &  [Fe/H]   &  ${\rm log}~g$  & $\Theta_{\rm{3D}}/\Theta_{\rm{1D}}$ & $\Delta \teff$ [K]\\
\hline
0.5 \footnote{central wavelength of the corresponding optical filter} & 0.0 &  2.2  & 1.003 & $-7$\\
2.14 \footnote{central wavelength of the corresponding FLUOR filter}& 0.0 & 2.2 & 0.996  & 9\\
\hline
0.5 & $-$1.0  &  2.2   & 0.991 & 21 \\
2.14& $-$1.0 & 2.2  &  0.996 & 9\\
\hline
0.5 & $-$2.0  &  2.2  & 0.982 & 46\\
2.14& $-$2.0 & 2.2  & 0.990 & 25 \\
\hline
0.5 & $-$3.0  &  2.2  & 1.011 & $-28$\\
2.14& $-$3.0 & 2.2  & 0.989  & 28\\
\hline
0.5 & $-$3.0  &  1.6  & 0.965 & 82 \\
2.14& $-$3.0 & 1.6 & 0.984 & 37 \\
\hline
\end{tabular}
\end{minipage}
\end{table}

These corrections could also be relevant in explaining some discrepancies known in 
literature, which involve the comparison with interferometric angular 
diameters. For example, in the library of stellar spectra assembled by 
\cite{1999AJ....117.1864C}, absolute spectrophotometry has the tendency to 
return angular diameters systematically smaller by a few percent with respect 
to interferometry. 
Smaller angular diameters for cool giants would also explain 
the dichotomy reported by \cite{1994A&A...282..684A} when trying to set on the 
same scale $\teff$ derived for hot and cool stars from interferometry with 
those determined from the Infrared Flux Method. Recent revision of the 
latter technique support in fact hotter effective temperatures also for dwarfs 
and subgiants \citep{2006MNRAS.373...13C,2010A&A...512A..54C} and the 
extension to giants is currently under way (Casagrande et al.~in prep.). 
Smaller limb-darkened angular diameters (i.e.~hotter effective temperatures) 
would also reduce the discrepancy recently noticed by 
\cite{2010ApJ...710.1365B} when comparing interferometric with spectroscopic 
$\teff$ of giants.
Thus, it is important to keep in mind that even when interferometry is used to 
derive effective temperatures in the most fundamental way, there is a non 
negligible --and still partly unexplored-- model dependence in the 
correction from uniform to limb-darkened disks. 
We remark that our results are currently limited to a rather small range in 
effective temperatures and gravities and to this respect, 3D models will also 
give an important contribution to set the effective temperature scale on a physically sounder base.

\subsection{Closure phases}

One can use closure phase between three telescopes, as the sum of all phase differences
removes the atmospheric contribution, leaving the phase information of
the object visibility unaltered \citep[e.g.][]{2007NewAR..51..604M}.
The closure phase thus offers an important complementary piece of information, revealing
asymmetries and inhomogeneities of stellar disk images.

\begin{figure}
  \centering
   \begin{tabular}{c}
     \includegraphics[width=1.0\hsize]{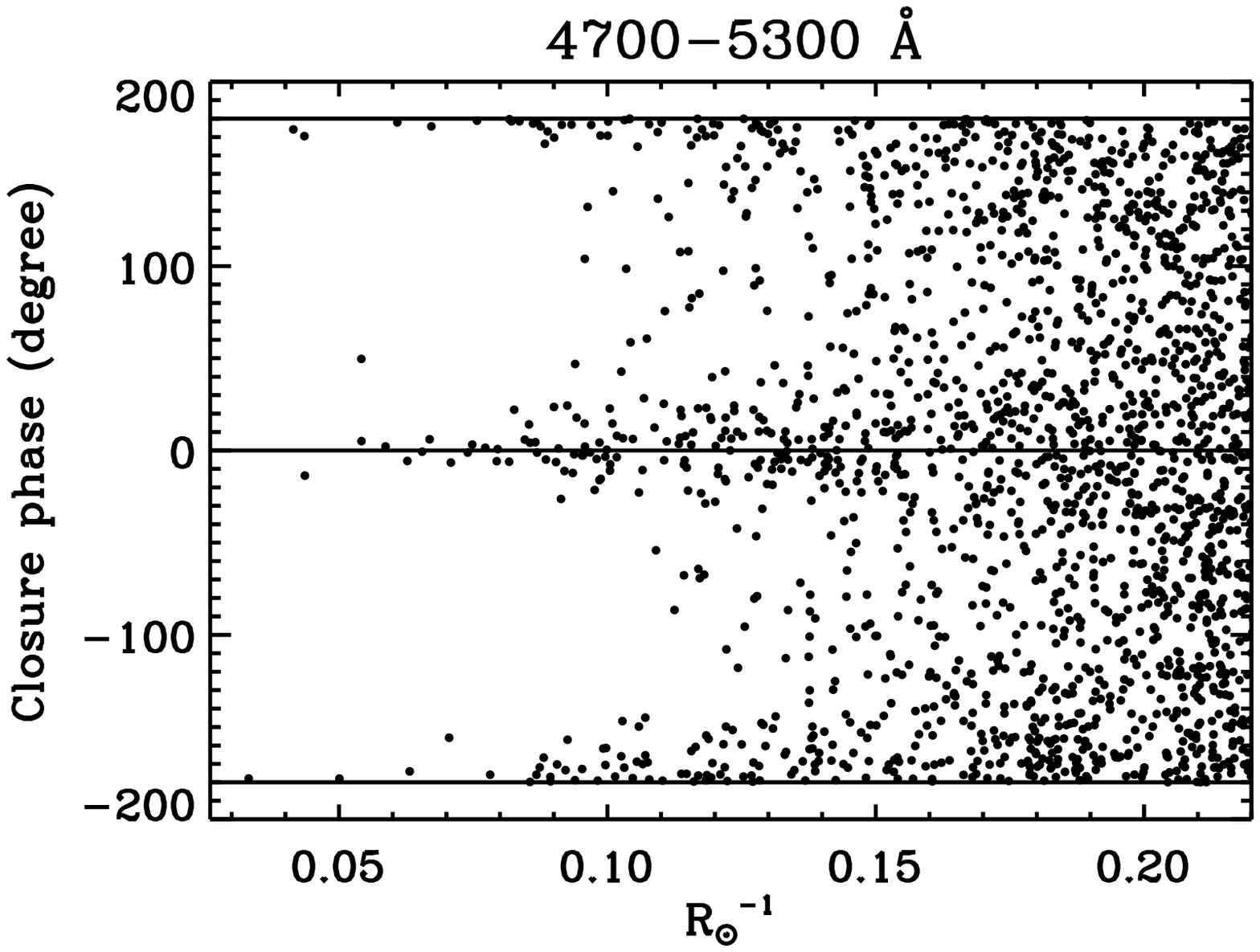} \\
 \includegraphics[width=1.0\hsize]{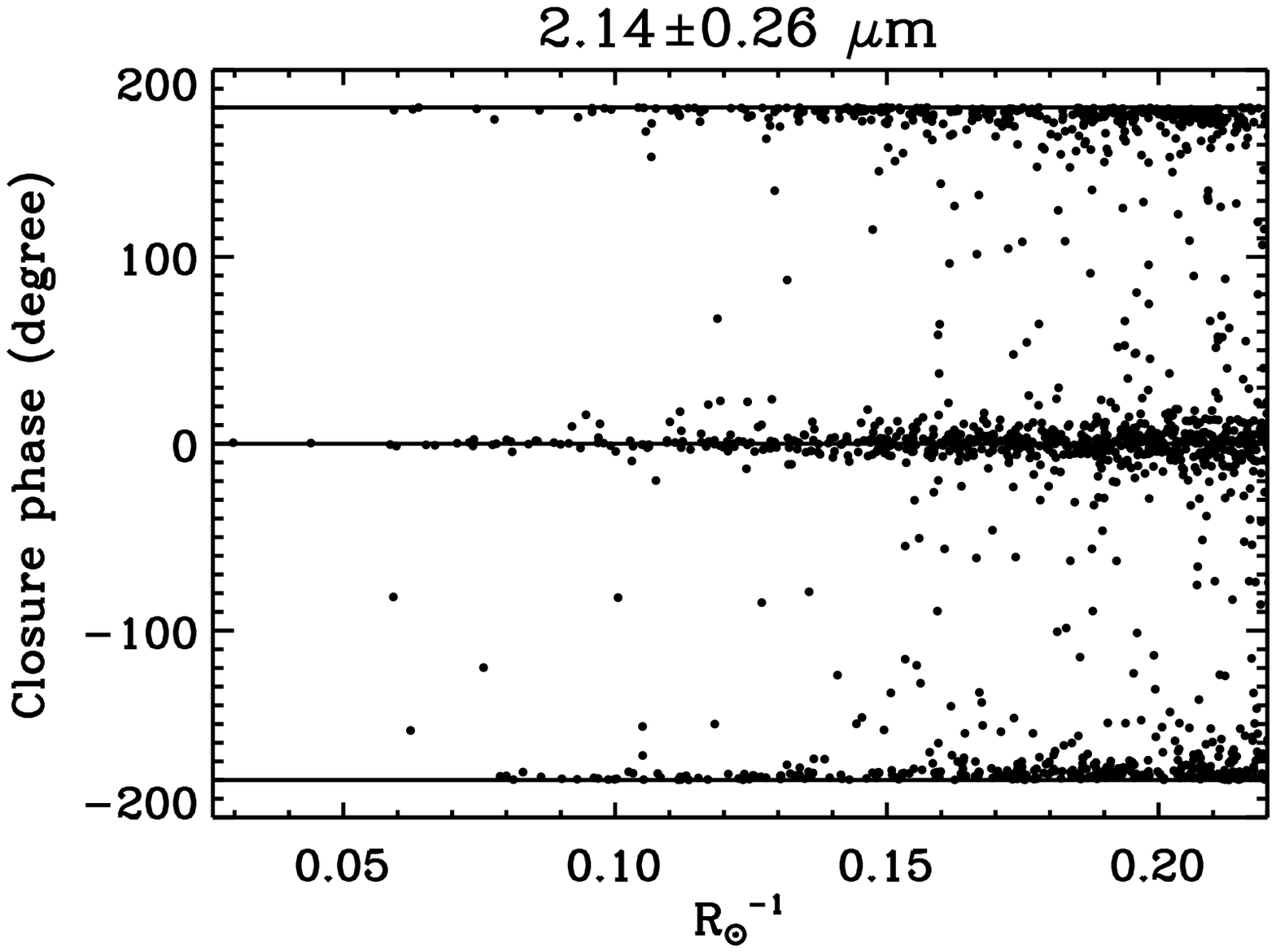} 
 \end{tabular}
     \caption{Scatter plot of closure phases of 2000 random baseline
       triangles for RHD simulation with
       [Fe/H]=$-$3.0 and $\log g$=1.6 with a maximum linear
       extension of $\sim$510 m at 0.5 $\mu$m and $\sim$2200 m at 2.14 $\mu$m and for
       a star at 100 pc. The extension of the baselines has been
       expressed in $\rm{R}^{-1}_\odot$ using Eq.~\ref{eqvis1}
       and~\ref{eqvis2}. The axisymmetric case is represented by the
       straigth lines.
          }
       \label{closure}
  \end{figure}

Figure~\ref{closure} displays large deviations from axisymmetric case (zero or $\pm\pi$). This is an evident signature of surface inhomogenities. There is a correlation between Fig.~\ref{closure} and Fig.~\ref{vis-fluct} because the scatter of closure phases increase with spatial frequencies as for the visibilities. Moreover, the optical filter shows larger deviations than the FLUOR one, and in both cases there are asymmetries already detectable at frequencies corresponding to the top of the second lobe. 
It would be very efficient to constrain the level of asymmetry and
inhomogeneity of stellar disk images by accumulating statistics on
closure phase at short and long baselines. Small deviations from zero
will immediately reveal departure from symmetry thus complementing
information of the visibility curves towards a better characterization of limb darkening and granulation patterns. The CHARA interferometer is well adapted for this purpose because it can combine several telescope at once and get reliable closure phases.

The characterization of the closure phase,
together with the limb darkening behavior (see Section~\ref{Sectlimb}), is important for two more reasons in red
giant stars. First, \cite{2008SPIE.7013E..91R}
demonstrated that closure phases provide the differential planet to
star contrast ratio as a function of wavelength. When observing a star
with a faint companion, their fringe patterns add up 
incoherently and the presence of a planet causes a slight decrease in
the phase changes and, consequently, the closure phases. This difference can be measured with a temporal survey and should be corrected with the intrinsic closure phases of the parental star. Second,
\cite{2008PASP..120..617V} provided a technique based on closure
phases to determine the 
orbital plane position angle of a planet transiting in front of the star as well as the planet's radius
using the high-precision multi-telescope beam combiner (MIRC)
\citep{2006SPIE.6268E..55M} on CHARA. Also in this case, theoretical
predictions of the closure phases of the parental stars are crucial.

\section{Comparison with the interferometric observations of HD~214868}\label{Sect:comparison}

\begin{figure}
  \centering
   \begin{tabular}{c}
     \includegraphics[width=1.0\hsize]{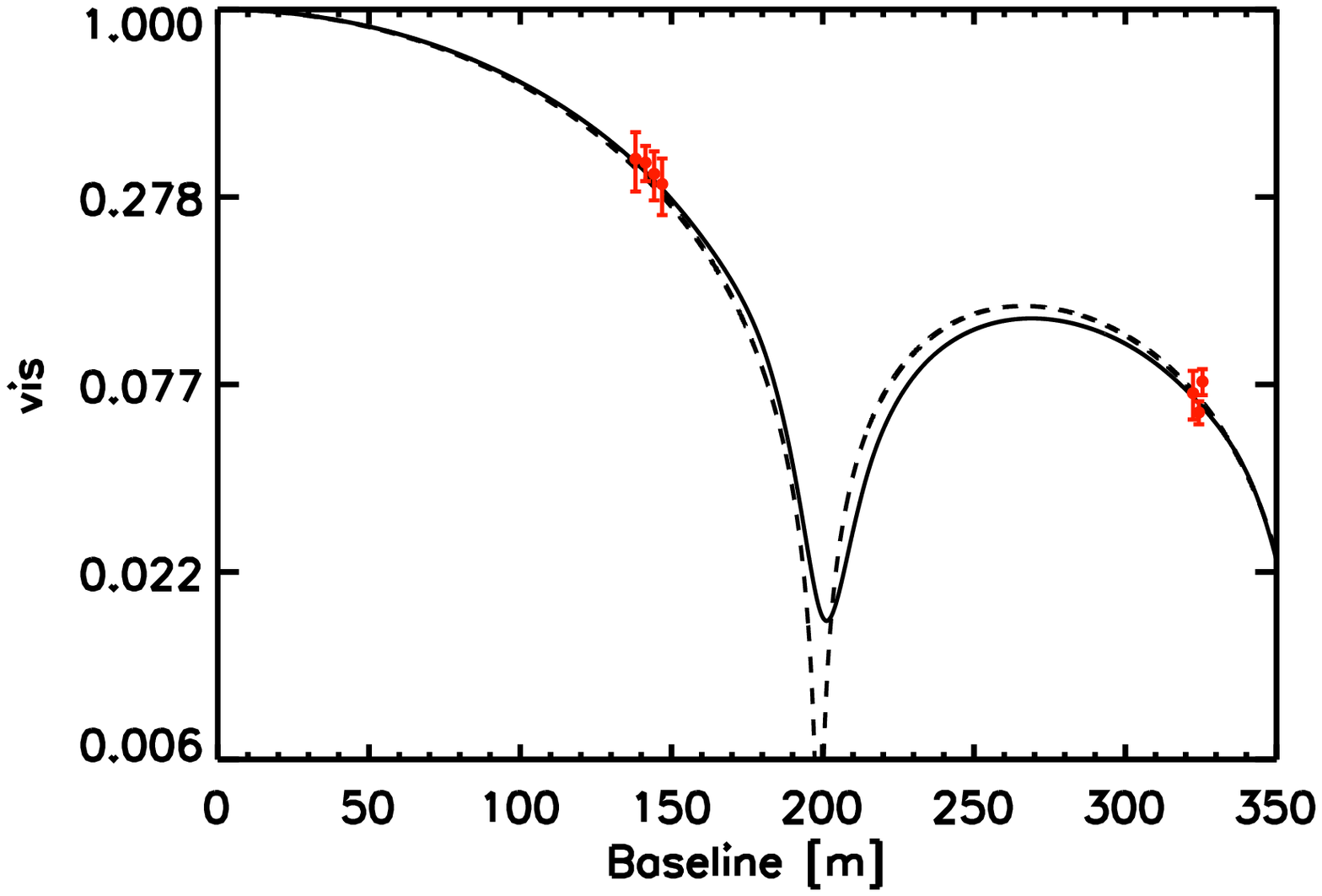} \\
 \includegraphics[width=1.0\hsize]{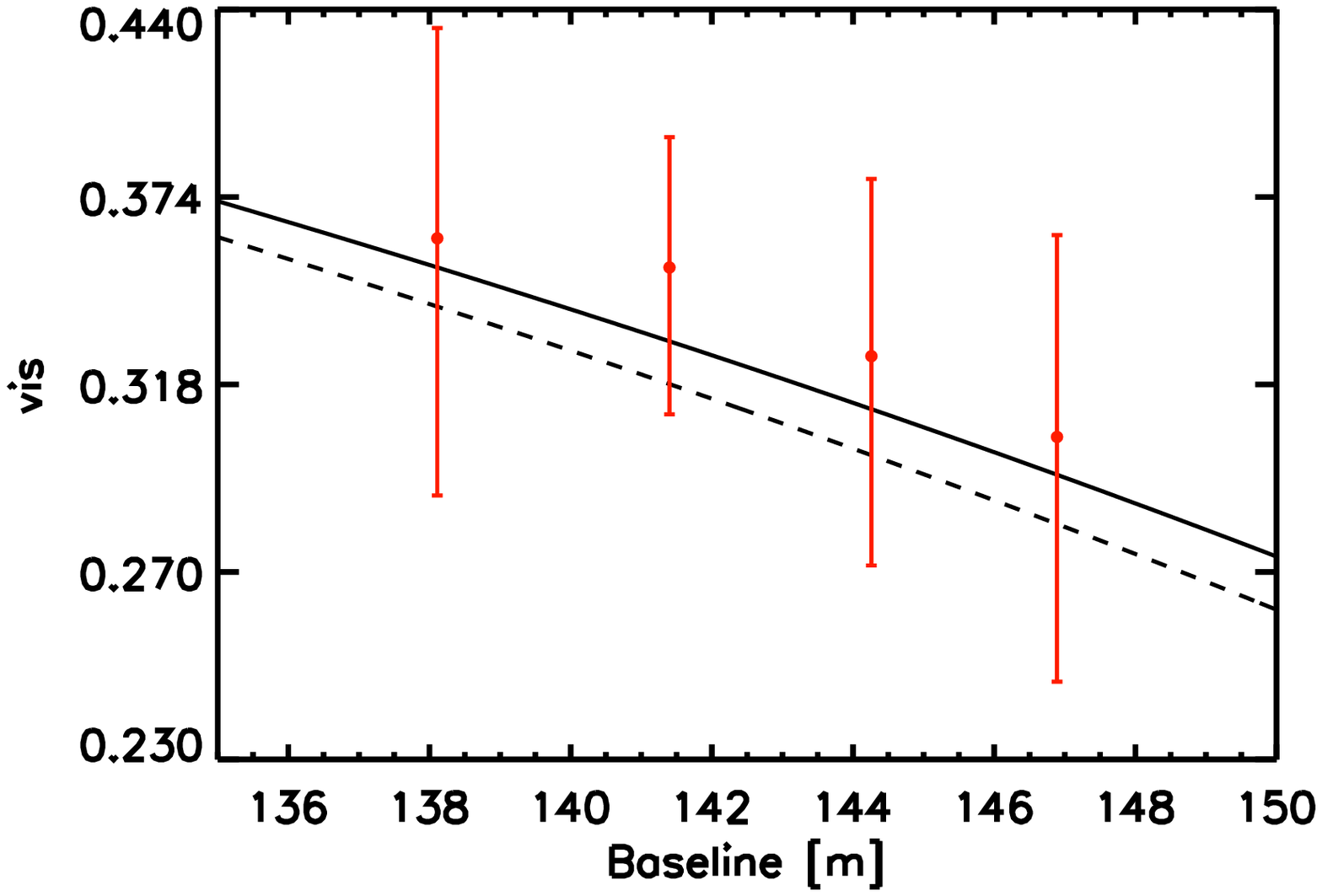} \\
  \includegraphics[width=1.0\hsize]{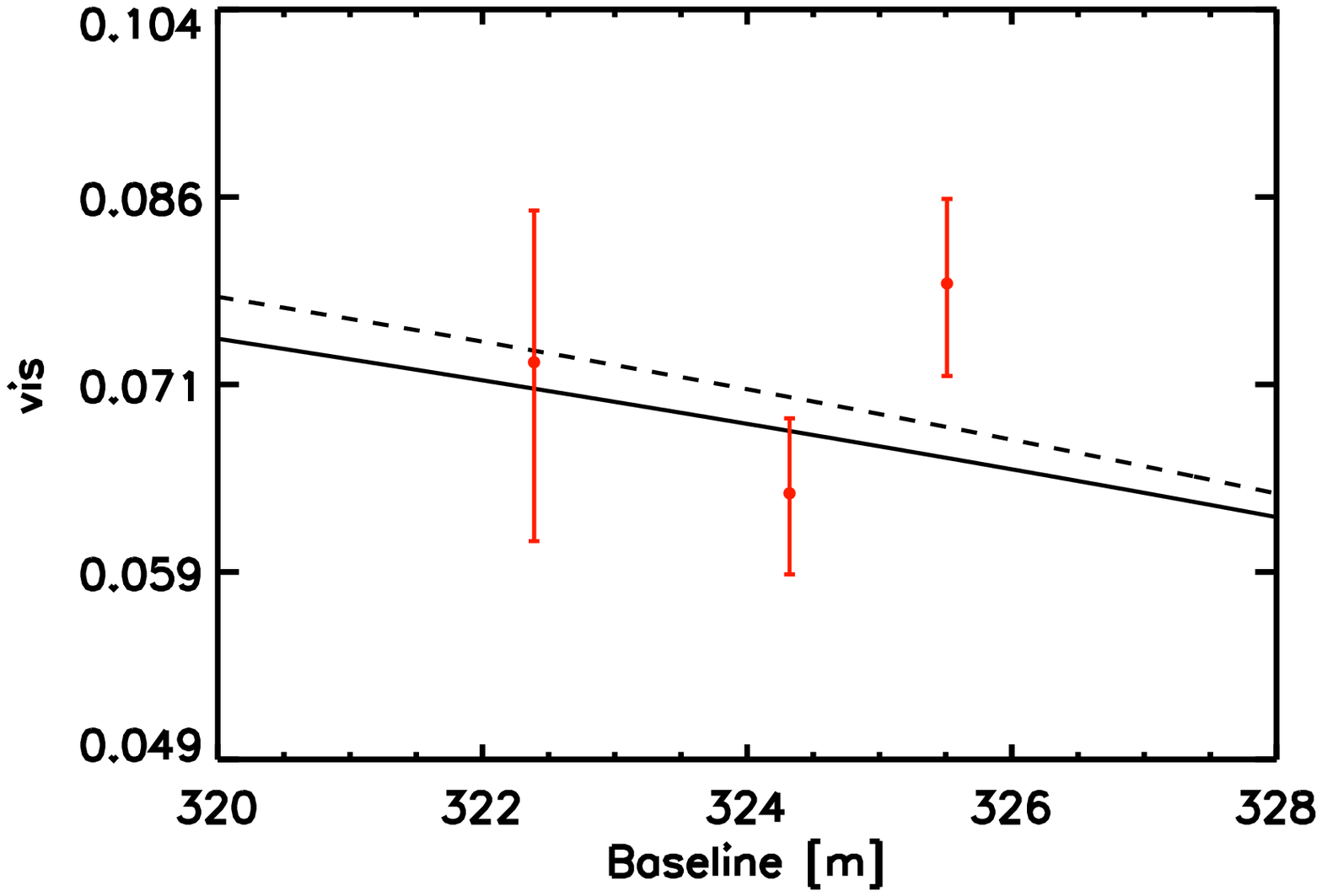} 
 \end{tabular}
     \caption{Comparison of the RHD simulation with [Fe/H]=0. and
       ${\rm log}~g$=2.2 (Table~\ref{simus}) to the observations (red
       dots) of the giant HD~214868 \citep{2010ApJ...710.1365B}. The
       solid thick line corresponds to the best fit (reduced
       $\chi^2$=0.59) visibility curve, the dashed line is the
       uniform disk (2.721 mas with reduced $\chi^2$=0.72) model
       introduced by \citeauthor{2010ApJ...710.1365B}. The visibility curve from RHD simulation never
goes to zero due to the bandwidth smearing effect. A logarithm
       scale is used on y-axis. 
          }
       \label{comparison}
  \end{figure}

In this Section, we compare our RHD simulations to the
giant HD~214868 \citep[K2~III, $T_{\rm{eff}}=4440\pm70$, $\log
 g=2.1\pm0.2$ and {[}Fe/H{]}= $-0.18\pm0.50$
 dex;][]{2010ApJ...710.1365B}. This star has been chosen because it
has been observed up to the second lobe with CHARA in a filter centered at
$\sim2.15$ $\mu$m and sensitive to the range $1.00-2.50$ $\mu$m
\citep{2005ApJ...628..439M}, which is close to
FLUOR filter. For this purpose, we chose the
model of Table~\ref{simus} having the closest parameters
(i.e. $T_{\rm{eff}}=4697\pm18$, [Fe/H]=0.0, and ${\rm log}~g$=2.2) and
compared our synthetic visibilities in the FLUOR filter to the observations (Fig.~\ref{comparison}). The absolute model
dimensions have been scaled to match the interferometric observation
in the first lobe. However, as already mentioned in
Section~\ref{SectVis}, the information about the radius cannot be 
retrieved because of the limitations in the method we use to construct the stellar disk images. We therefore concentrate on the visibility data points
in the second lobe \citep[i.e. where the limb darkening effects are
 important;][]{1974MNRAS.167..475H}. A Levenberg-Marquardt
least-squares minimization returns reduced
$\chi^2$=0.59 for the best RHD synthetic visibility, all RHD
visibilities falling in the range [0.59, 0.66]. Formally, the RHD
simulation shows an improvement over parametric model (uniform disk model has
reduced $\chi^2$=0.72) for the interpretation of these interferometric
observations, though it must be noted that observed points in
Fig.~\ref{comparison} fall at baselines where 3D and uniform disk
models nearly overlap. A more uniform sampling (e.g. close to the
first null point) will be very valuable to further test our
simulation. For the sake of completeness, we also remark that the
simulation used in this specific exercise is not tailored exactly to
the stellar parameters of the studied star, and improvements along
this direction are still possible.

\section{Conclusions}

In this work, we provided a set of interferometric predictions from our RHD simulations of red giant stars that will be tested against observations with today interferometers like CHARA. Red giant stars are ideal targets for interferometers because they are numerous and have large angular diameters as well as bright infrared apparent magnitudes.

We derived average center-to-limb intensity profiles from the
synthetic stellar disk images based on RHD simulations and computed limb-darkening 
coefficients in the optical as well as in the infrared. The
predicted center-to-limb variations of 3D simulations can be tested
against observations: the visibilities and the closure phases show
evident departures from from circular symmetry, which are
due to inhomogeneities on the stellar surface. We emphasize that
red giant stars should be observed at high spatial frequencies, to
characterize the granulation pattern, and in
particular in the optical range where the visibility fluctuations
are larger. In addition to this, it would be valuable to couple the observations in the optical and in the infrared to test models predictions
for limb darkening. We counted at least 58 red giants stars using the stellar diameters catalogue of \cite{2010yCat.2300....0L} with stellar parameters corresponding to our simulations, an apparent diameter larger than 2 mas in the visible and $V$ magnitude lower than 6 \citep[the current magnitude limit of VEGA is 7,][]{2009A&A...508.1073M}. We provided also a first comparison
with the red giant star HD~214868 showing that our RHD simulations
returns a better fit with respect to simple parametric uniform disk
models. More observations along this direction will provide useful
information not only on red giants, but also on our RHD models. This will include the
center-to-limb variation, which carry the information on the
temperature stratification in the stellar atmosphere, and the
characterization of the granulation pattern in terms of size and
contrast with varying metallicity and effective temperature. Moreover,
the study of the closure phases behavior will help to recover the
physical parameters of possible hot Jupiters around red giant stars.

Interferometry is advantageous in that it provides the ability to directly
measure stellar angular diameters. We found that, for typical red
giant stars studied in this work, differences in angular diameters
using 3D or 1D limb-darkening laws
vary from $\sim -3.5\%$ to $\sim 1\%$ in the optical, and are roughly in the 
range between
$-0.5$ and $-1.5\%$ in the infrared, with 3D models returning in general smaller angular diameters with respect to 1D, thus 
implying moderately hotter effective temperatures. The impact of these
corrections is not dramatic, but they are not negligible to correctly
set the zero point of the effective temperature scale derived by means of this 
fundamental method, in particular, for future observations of
metal-poor stars. Moreover, the model dependence on the diameter
determination has also an impact on the reliability of the existing
catalogues of calibrator stars for interferometry that are based on
giants with parameters similar to those covered in this work. 
We emphasize that even when interferometry is used to 
derive effective temperatures in the most fundamental way, there is a
non-negligible and still partly unexplored model dependence.

\begin{acknowledgements}
We thank the Rechenzentrum Garching (RZG) for providing some of the computational resources necessary for this
work.  
\end{acknowledgements}

\bibliographystyle{aa}
\bibliography{biblio.bib}

\clearpage

\end{document}